\documentclass[]{gji}
\usepackage[english]{babel}
\usepackage[T1]{fontenc}
\usepackage{timet,color}
\usepackage[urlcolor=blue,citecolor=black,linkcolor=black]{hyperref}

\usepackage{graphicx}
\usepackage{amsmath}
\usepackage{amssymb}
\usepackage{indentfirst}
\usepackage{apalike}
\let\bibhang\relax
\usepackage{natbib}
\bibpunct[, ]{(}{)}{,}{a}{}{,}%
\def\bibfont{\small}%
\def\bibsep{\smallskipamount}%
\def\bibhang{24pt}%
\title[Velocity-spreading factor and consequences]{Exploring velocity-spreading factor and consequences through dynamic ray-tracing in general anisotropic media: A comprehensive tutorial}

\author[Coimbra et al]{Tiago A. Coimbra$^1$, Rodrigo Bloot$^2$, and Jorge H. Faccipieri$^1$ \\
$^1$ University of Campinas, Center for Energy and Petroleum Studies, High-Performance Geophysics Lab\\
Rua Cora Coralina, 350, Cidade Universit\'aria, 13089-970, Campinas, SP, Brazil \\
$^2$ Federal University of Latin America Integration, \\
Avenida Silvio Americo Sasdelli,  6731 - Bloco 4, 85867-970, Foz do Igua\c{c}u, PR, Brazil
}
  
\begin{document}

\label{firstpage}

\maketitle

\begin{abstract}
In seismic imaging, understanding the relationship between wavefront-propagation velocity and time-interval velocity is crucial for achieving optimal resolution. However, this task becomes even more challenging when considering anisotropic situations. To accurately account for the influence of anisotropy on wavefronts, it is essential to have a solid grasp of the underlying physics. Unfortunately, the anisotropy model that best describes the medium is often unknown. To address this issue, we utilize paraxial-ray theory in a ray-centered coordinate system to study the wavefront phenomenon. This approach allows us to develop explicit expressions that describe the physics of the problem. Using this theoretical framework, we can accurately generalize the relationship between time-migration rays and Dix velocity by incorporating the velocity-spreading factor for general anisotropic media. This factor lets us determine the type of anisotropy present in the medium. Moreover, the velocity-spreading factor provides valuable information for various applications, including model building, time-imaging, and time-to-depth conversion. Overall, the presented theoretical framework offers a comprehensive understanding of wavefront propagation in anisotropic media, which can aid in improving the knowledge of the phenomena that form seismic images.
\end{abstract}


\section{Introduction}

The presence of anisotropy in the medium imposes additional difficulties in seismic processing. Practical frameworks have been presented in literature and applied to models with anisotropy concerning a medium axis of symmetry, e.g., transversely isotropic media, because of its intrinsic less complex mathematical expressions and several real cases for applicability \citep[e.g.,][]{Grechka:1996, Grechka:1998, Sadri:2010, Tohti:2021, Sethi:2021}. In his pioneering work \cite{Thomsen:1986} presented a formulation with practical attributes to describe the physics of the problem representing the anisotropy by simple parameters. Historically, \cite{Alkhalifah:1995} and \cite{Tsvankin:2001}, for example, incorporate those parameters for seismic processing. They were very successful in their formulation by making easy, for purposes of the processing, the understanding of the influence of the anisotropy parameters on stacking velocities for a locally homogeneous medium. Therefore, to avoid locally homogeneous approximation approaches, we described the anisotropy factor as the parameter related to the deviation presented by the phase and group velocity. The main focus is to characterize phase velocity variation in direction (anisotropy) and position (heterogeneity) to quantify how the anisotropy deviation factor influences such attributes for any anisotropy. For this characterization to be adequate, it is necessary to correctly describe the wavefront phenomenon in a specific coordinate system. Based on this premise, the entire development of the main text is based on the explicit formulation in terms of phase velocity and its first- and second-order derivatives in such a coordinate system.

In order to study the complex behavior of seismic wavefronts, it is convenient to use the dynamic ray theory. Such a theoretical approach is well consolidated for the isotropic media with many practical applications in the literature \citep[e.g.,][]{Popov:1978,Farra:1987,Cerveny:2001,Popov:2002,Cameron:2007,Iversen:2008}. However, concerning anisotropic media in general, there is still a lack of results to be explored. Despite such lack, several authors have studied this problem in many publications \citep[e.g.,][]{Cerveny:1972,Cerveny:1985,Cerveny:2001,Klime1994,Cerveny:2009,Cerveny:2010,Iversen:2018,Iversen:2021}.

Despite being crucial to many applications, the ray theory imposes several challenges for an understandable description of the physical problems. Also, its mathematical modeling can be complex for general anisotropic media. Therefore, in the context of this work, we use paraxial-ray theory \citep{Popov:2002} to study the ray theory phenomena in anisotropic media through a straightforward process, introducing explicit expressions for the Lagrangian and Hamiltonian for general anisotropic heterogeneous media. Such an approach allows the application of mathematical terms to understand the physical features described by the model and the construction of algorithms for applications accessible without necessarily a deep background about the theme. In other words, it offers a theoretical procedure with adequate formalism and applicability.

Understanding the physical phenomenon through straightforward mathematical expressions helps to generalize isotropic case well-established results to anisotropic media. 
Besides, a formal description of the wavefront propagation phenomenon also provides relevant tools for seismic processing and interpretation. Based on these premises, we present a physical interpretation of the emerging wavefronts \citep{Hubral:1980} measured at the surface and how anisotropy influences its shape. We generalize the results obtained by \cite{Cameron:2007} concerning the velocity-spreading factor effect for general anisotropic media. In this way, we show that the spreading factor is related to heterogeneity and the anisotropy of the medium. Besides, we elucidate the natural connection between the Dix velocity \citep{Dix:1955} and its relation to the time-migration wavefront. Consequently, the results obtained by these studies have immediate application for a precise understanding of the time-migration ray \citep{Fomel:2021}. From these studies, we obtain a relationship between Dix and group velocities, which helps to determine what type and degree of anisotropy exist in the medium. Finally, as already mentioned, the explicit expressions of the Hamiltonian are essential to understanding the physical phenomenon and allow the introduction of easy-to-implement algorithms for processes such as modeling, imaging, and time-to-depth conversion.

In the following sections, we provide a comprehensive and sequential development of our formulation with a brief description in the following. In the ray tracing in the Cartesian coordinate system section, we revisit the ray theory to formulate the problem through the phase velocity approach from the elastodynamic wave equation. We show that developing this problem as such has particular advantages compared to the usual procedure from the literature \cite[see, e.g.,][]{Aki:Richards:1980,Cerveny:2001,Popov:2002,Pujol:2003}. In the formulation in ray-centered coordinates section, we expand results related to the ray theory in a non-orthogonal-centered coordinate system to improve our understanding of the physical phenomenon in place. Besides, we present Lagrangian and Hamiltonian formulations of the paraxial-ray theory in that coordinates system. In the dynamic ray tracing system section, we use the Hamiltonian system to analyze the dynamic properties along the central ray using a non-orthogonal system to reduce the system equations. Besides, we show the elements of the propagator matrix in our formulation together with the initial conditions for each type of wavefront based on starting point. In the physical interpretation of the emerging wavefronts and application section, we provide interpretations of the physical attributes of the wavefront measured at the surface and obtain the exact normal-moveout (NMO) velocity derived directly from the proposed formulation. In the velocity spreading factor in general anisotropic media section, we present the relationship between the Dix velocity and the time-migration curvature and extend the work of \cite{Cameron:2007} to general anisotropic media. Finally, in the Eikonal-type equation for time migration section, we derive an Eikonal-type equation for time migration, which determines the time-migration rays \citep{Fomel:2021} and show a relation between Dix and time-migration ray velocities and, therefore, we describe a way to determine the anisotropy of a medium.

\section{Ray tracing in Cartesian coordinate system}

This section describes, as a tutorial, a ray-tracing algorithm in Cartesian coordinates representing an approximate high-frequency solution for the elastodynamic wave equation. This solution leads the wavefield decomposition into independent contributions called elementary waves, propagating along the raypaths, representing seismic body waves propagating in a medium. Such decomposition provides a significant advantage that allows for separate analysis and handling of each individual-elementary wave. Besides, we briefly review the construction steps to obtain an expression in the function of phase velocity of the Hamiltonian system in Cartesian coordinates in anisotropic media.

Considering the absence of external forces, such as body forces, we have that the general wave equation \citep[e.g.,][]{Aki:Richards:1980} for heterogeneous anisotropic media in the frequency-domain is given by
\begin{equation}
    -\omega^2\rho u_i=\sum_{j=1}^{3}\sum_{\text{k}=1}^{3}\sum_{l=1}^{3}\frac{\partial}{\partial x_j}\left(c_{ijkl}\frac{\partial u_{k}}{\partial x_{l}}\right)\, , \quad i = 1,2,3\, ,
    \label{geral_aniso}
\end{equation}
where $u_{i} = u_{i}(\bar{\bf x};\omega)$ represents the elements of the displacement vector. We denote $\bar{\mathbf{u}}=[u_1,u_2,u_3]^T$, where the upper-scale letter $T$ represents the transpose operator, with 3D Cartesian coordinates $\bar{\mathbf{x}}=[x_1,x_2,x_3]^T$, and $\omega$ as the circular frequency. Heterogeneity and anisotropy are embodied by the stiffness tensor $c_{ijkl} = c_{ijkl}(\bar{\bf x})$ which, depending on the medium, has several symmetries that allows for simplifications on its expression \citep[see, e.g.,][ for more details]{Aki:Richards:1980,Cerveny:2001,Pujol:2003} and $\rho=\rho(\bar{\bf x})$ is the time-independent density parameter.

In order to obtain an asymptotic solution, we take the ray-ansatz solution \citep{Sommerfeld:1911} as follows
\begin{equation}
\bar{\bf u}(\bar{\bf x},\omega) = \bar{\bf u}_0(\bar{\bf x}) e^{-{\cal I}\omega\tau(\bar{\bf x})}\, ,
\label{eq:ansatz}
\end{equation}
where $\bar{\bf u}_0(\bar{\bf x})$ is the polarization vector, ${\cal I}$ is the unit imaginary number, and $\tau(\bar{\bf x})$ is the traveltime function. Therefore, substituting eq.~(\ref{eq:ansatz}) in eq.~(\ref{geral_aniso}), and taking as null the term multiplied by $\omega^2$, we obtain the Christoffel equation \citep{Christoffel:1877}
\begin{equation}
\bar{\bf W}\bar{\bf u}_0 = \bar{\bf u}_0\, ,
\label{autovetor}
\end{equation}
with the components of the Christoffel matrix, $\bar{\bf W} = [W_\text{ik}]$, being
\begin{equation}
W_{ik}(\bar{\bf x},\bar{\bf w}) = \sum_{j=1}^{3}\sum_{l=1}^{3} a_{ijkl}(\bar{\bf x})w_{j} w_{l}\, ,
\label{eq:normal:Gamma}
\end{equation}
where $a_\text{ijkl} = c_\text{ijkl}(\bar{\bf x})/\rho(\bar{\bf x})$ is the density-normalized elastic moduli, and $\bar{{\bf w}}=[w_1, w_2, w_3]^T$ represents the slowness vector, i.e.,
\begin{equation}
\bar{\bf w} = \frac{\partial\tau}{\partial \bar{\bf x}} = \left[\frac{\partial\tau}{\partial x_1},\frac{\partial\tau}{\partial x_2},\frac{\partial\tau}{\partial x_3}\right]^T\, .
\label{eq:teqtau}
\end{equation}

In eq.~(\ref{autovetor}), we observe that the polarization vector $\bar{\bf u}_0$ is an eigenvector of $\bar{\bf W}$ with eigenvalue equal to one. Also, from \cite{Cerveny:2001}, the Christoffel matrix $\bar{\bf W}$ has at least three more important properties. First, $\bar{\bf W}$ is symmetric, i.e., $W_{ik} = W_{ki}$. Second, the elements $W_{ik}$ are homogeneous functions of the second degree in $\bar{\bf w}$, i.e., $W_{ik}(\bar{\bf x},v\bar{\bf w}) = v^2W_{ik}(\bar{\bf x},\bar{\bf w})$. Lastly, $\bar{\bf W}$ is positive definite, i.e., $\hat{\bf n}\cdot\bar{\bf W}\hat{\bf n} > 0$ for any unit vector $\hat{\bf n}$, where the symbol $\cdot$ represents the inner-product operator. Thus, taking the unit vector $\hat{\bf n}$ from $\bar{\bf w}$, as $\hat{{\bf n}}=[n_1, n_2, n_3]^T$, we have Christoffel's matrix in its normalized form, namely,
\begin{equation}
N_{ik}(\bar{\bf x},\hat{\bf n}) = \sum_{\text{j}=1}^{3}\sum_{\text{l}=1}^{3} a_{ijkl}(\bar{\bf x})n_{j} n_{l}\, .
\label{eq:normal:N}
\end{equation}
Moreover, the value $v(\bar{{\bf x}};\hat{{\bf n}})$ is the phase velocity of the respective wave mode, which over the raypath has the following form
\begin{equation}
\bar{\bf w} = \|\bar{\bf w}\|\hat{\bf n} = \frac{\hat{\bf n}}{v}\, .
\label{eq:raypath}
\end{equation}
Besides, from the property of homogeneous function, we can observe the following equalities
\begin{equation}
v^2\bar{\bf W}(\bar{\bf x},\bar{\bf w})=\bar{\bf W}(\bar{\bf x},v\bar{\bf w})=\bar{\bf N}(\bar{\bf x},\hat{\bf n})\, .
\label{eq:WN}
\end{equation}
Based on symmetry and positive definite property about $\bar{\bf W}$ and by eq.~(\ref{eq:WN}), the normalized Christoffel matrix $\bar{\bf N}$ has three positive real eigenvalues, implying three corresponding mutually orthogonal eigenvectors. We denote the eigenvalues of $\bar{\bf N}=[N_{ik}]$ by the symbol $N_\lambda(\bar{\bf x}, \hat{\bf n}) = v^{2}(\bar{\bf x}; \hat{\bf n})$, in which the values of $\lambda$ correspond to P, SV, or SH wave modes, i.e., the values $N_\lambda$ are the roots of the characteristic polynomial of $\bar{\bf N}$ described as 
\begin{equation}
\det[\bar{\bf N} - v^2\bar{\bf I}] = (v^2-N_{P})(v^2-N_{SV})(v^2-N_{SH})\, .    
\end{equation}
It is important to clarify that in our approach, the phase velocity is a function of the normal vector and not the slowness vector due to the phase velocity squared to be eigenvalues of $\bar{\bf N}$. From the physical point of view, taking the matrix eigenvalues as phase velocity allows us to directly analyze the influence of anisotropy along its rates of change. From the mathematical point of view, as this formulation allows using the unit vector of the phase, this makes the formula not recursive concerning the phase velocity itself. Such a formulation also allows us to derive general forms of the equations governing ray theory in anisotropic inhomogeneous media, namely, the Hamilton-Jacobi equation \citep{Gelfand:2000}.

Now, considering all previous assumptions, from eqs.~(\ref{autovetor}) and~(\ref{eq:normal:N}), an admissible phase vector $\bar{\bf w}$ is such as that
\begin{equation}
\bar{\bf w}\cdot\bar{\bf N}\bar{\bf w} = \hat{\bf n}\cdot\bar{\bf W}\hat{\bf n} = 1\, ,  
\label{norm}
\end{equation}
if, and only if
\begin{equation}
\bar{\bf w}\cdot\left(\bar{\bf N}-v^{2}(\bar{\bf x},\hat{\bf n})\bar{\bf I}\right)\bar{\bf w} = 0\,.
\label{norm1}
\end{equation}
As a consequence, directly from eqs.~(\ref{norm}) and~(\ref{norm1}) we have the following expression
\begin{equation}
    v^{2}\left(\bar{{\bf x}},\frac{\bar{{\bf w}}}{\|\bar{{\bf w}}\|}\right)\|\bar{\bf w}\|^2 = 1.
\end{equation}
Therefore, we can take a Hamiltonian that describes the kinematics of wave motion in Cartesian coordinates such as ${\cal H}_t(\bar{{\bf x}}, \bar{{\bf w}})$, which leads to
\begin{equation}
{\cal H}_t(\bar{{\bf x}},\bar{{\bf w}}) = \frac{1}{2}v^{2}\left(\bar{{\bf x}},\hat{\bf n}(\bar{\bf w})\right)\|\bar{\bf w}\|^2 = \frac{1}{2}\, .
\label{eq:Eikonal}
\end{equation}

The Hamiltonian, eq.~(\ref{eq:Eikonal}), can be solved using the method of characteristics \citep{Courant:1989}. This method provides the characteristic trajectories along which propagation occurs from one start-point position to another. Besides, the trajectory of the Hamiltonian, eq.~(\ref{eq:Eikonal}), in global system is given by
\begin{equation}
\frac{d\bar{\bf x}}{dt} = \frac{\partial {\cal H}_t}{\partial\bar{\bf w}}\, , \quad
\frac{d\bar{\bf w}}{dt} = -\frac{\partial {\cal H}_t}{\partial\bar{\bf x}} \, .
\label{system1}
\end{equation}
Furthermore, as in seismic literature \citep[e.g.,][]{Bleistein:1984}, we describe the ray method as the solution of these characteristic trajectories, which also provides a natural synthesis of mathematical and physical insights into wave propagation. Mathematically, the ray methods extend partial differential equation problems through the ray-anzats approach in an ordinary differential equation problem \citep{John1971}. Physically, ray methods develop the basic concepts of geometrical optics to a large class of optical wave phenomena and then extend these results to other wave phenomena \citep{Pujol:2003}.

In order to make an analysis of eq.~(\ref{system1}),
we assume a wavefront moving through space can be represented by $t - \tau(\bar{\bf x}) = 0$, based on the notation in eq.~(\ref{eq:teqtau}), which implies
\begin{equation}
dt = \frac{\partial \tau}{\partial \bar{\bf x}}\cdot d\bar{\bf x} = \bar{\bf w}\cdot d\bar{\bf x}\, .
\end{equation}
Therefore, one can show that the left equation of eq.~(\ref{system1}) yields
\begin{equation}
\bar{\bf w}\cdot\frac{\partial {\cal H}_t}{\partial\bar{\bf w}} = \bar{\bf w}\cdot \frac{d\bar{\bf x}}{dt} = 1\, .
\label{eq:relation_pdx}
\end{equation}
Considering an abuse of notation to make it easier to read, we assume
\begin{equation}
\frac{\partial v}{\partial \bar{\bf w}} = \frac{\partial \hat{\bf n}^T}{\partial \bar{\bf w}}\frac{\partial v}{\partial \hat{\bf n}}\, . \end{equation}
Besides, the first vector-form equation set in eq.~(\ref{system1}) describes the group velocity vector, which is tangent to the ray. Therefore, applying such a formulation into eq.~(\ref{eq:Eikonal}) and together with property in eq.~(\ref{eq:relation_pdx}), we have
\begin{equation}
 \frac{d\bar{\bf x}}{dt}= v^2\bar{\bf w} + \frac{1}{v}\frac{\partial v}{\partial \bar{\bf w}}\, , 
\label{grupofase}
\end{equation}
which implies, on the raypath, the following information
\begin{equation}
 \bar{\bf w}\cdot\frac{\partial v}{\partial \bar{\bf w}} = 0\, .
\label{ortogonal}
\end{equation}
In other terms, the phase velocity variation vector is always orthogonal to the phase vector because $v$ is homogeneous of degree zero in $\bar{\bf w}$. Given the previous observations, it is essential to mention that any approximation of the phase velocity must obey eq.~(\ref{ortogonal}) relation for the situation given by eq.~(\ref{eq:relation_pdx}) to occur.

In the literature \citep[e.g.,][]{Helbig:1994}, it is known that the group velocity vector is not necessarily perpendicular to the wavefront. One consequence is that the velocity to be analyzed concerning the wave propagation kinematics is the phase velocity. Therefore, taking the group velocity module, $v_g$, by
\begin{equation}
 v_g = \left\|\frac{d\bar{\bf x}}{dt}\right\|\, ,
\label{groupvel}
\end{equation}
together with the cosine law in eq.~(\ref{eq:relation_pdx}), yields \begin{equation}
v_g = \frac{v}{\cos\psi}\, ,
\label{gruprel2}
\end{equation}
with
\begin{equation}
\frac{1}{\cos\psi} = \sqrt{1 + \frac{1}{v^4}\left\|\frac{\partial v}{\partial \bar{\bf w}}\right\|^2}\, .
\label{eq:cosseno}
\end{equation}
Consequently, the proposed approach gets the same result presented in \cite{Tsvankin:2001}, for example. Therefore, the present formulation based only on phase velocity is consistent with the literature.

\begin{figure*}
\includegraphics[width=0.65\textwidth]{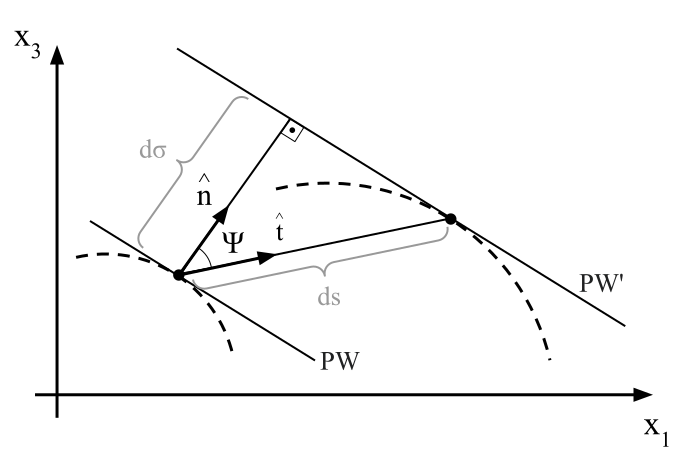}
\caption{Difference between the wavefront propagation related to the group velocity, represented by the dashed arcs, and the plane wave propagation, $PW$, that is related to the phase velocity. The angle $\psi$ represents the anisotropy deviation in the wave propagation phenomena.}
\label{fig:1}
\end{figure*}

The group velocity module, eq.~(\ref{groupvel}), which is related to the direction of the energy flow, is associated with the phase velocity through a $\psi$-angle deviation factor related to the medium anisotropy. However, this deviation factor does not occur when phase and group velocities coincide. The isotropic media represents one of such cases. In other words, regardless of the degree of anisotropy of the medium, the parameter $\psi$ provides a relationship between the group and phase velocities as seen in Figure~(\ref{fig:1}) represented by the angle between both velocity vectors and its variation quantifies the anisotropy deviation in non-dispersive media.
In order to show a feasible computational version for eq.~(\ref{system1}), by Appendix~\ref{Appendix:A}, the ray-tracing system in Cartesian coordinates given by eq.~(\ref{system1}) is described as
\begin{equation}
\frac{d\bar{\bf x}}{dt} = \frac{\partial v}{\partial\hat{\bf n}} + \left(v - \frac{\partial v}{\partial\hat{\bf n}}\cdot\hat{\bf n}\right)\hat{\bf n}\, , \quad \frac{d\bar{\bf w}}{dt} = -\frac{1}{v}\frac{\partial v}{\partial \bar{\bf x}}\, .
\label{eq:systemEDO}
\end{equation}
Furthermore, since $v^2$ is a homogeneous function of degree two in $\hat{\bf n}$, this implies that $v$ is a homogeneous function of degree one, that is,
\begin{equation}
\hat{\bf n}\cdot\frac{\partial v}{\partial\hat{\bf n}} = v\, ,   
\end{equation}
and using this result in eq.~(\ref{eq:systemEDO}), yields
\begin{equation}
\frac{d\bar{\bf x}}{dt} = \frac{\partial v}{\partial\hat{\bf n}}\, .  
\label{eq:dvdn_grupo}
\end{equation}
In other words, we now have described the group vector as the normal derivative of the phase velocity. We now analyze some particular cases to understand properties rendered by the parameter $\psi$. Furthermore, in this work, we define a medium at a given point $\bar{\bf x}$ as isotropic when its phase velocity at that point, $v(\bar{\bf x},\hat{\bf n})$, is constant when measured along all directions on a unit vector $\hat{\bf n}$, i.e., 
\begin{equation}
v^2(\bar{\bf x},\hat{\bf n}) = c^2(\bar{\bf x})\|\hat{\bf n}\|^2 \triangleq c^2(\bar{\bf x})\, .
\end{equation}
Otherwise, the medium has anisotropy at this point $\bar{\bf x}$. Note that $c(\bar{\bf x})$ represents the wave-propagating velocity for an isotropic medium depending on the position only, which characterizes the heterogeneity of the medium. Therefore, for an isotropic medium, we have the following results
\begin{equation}
\frac{1}{2}\frac{\partial v^2}{\partial\hat{\bf n}} = v\frac{\partial v}{\partial\hat{\bf n}} = c^2\hat{\bf n}\, ,    
\end{equation}
for all $\hat{\bf n}$. In addition, if there exists some situation such that
\begin{equation}
\frac{\partial v}{\partial\hat{\bf n}}= v\hat{\bf n}\, ,
\label{eq:vnw}
\end{equation}
we say there is an axis of symmetry of the anisotropic properties, which is parallel to $\hat{\bf n}$. Accordingly, by eq.~(\ref{eq:dvdw:dvdn}), we have
\begin{equation}
    \frac{\partial v}{\partial\bar{\bf w}}=\bar{\bf 0}\, .
    \label{ortonull}
\end{equation}
This implies that the group vector is the same as the phase vector; therefore, $\psi(\hat{\bf n})$ is null in such a symmetry axis. Besides, we say slowness is elliptical when the phase velocity satisfies \citep{Burridge:1993}
\begin{equation}
v(\bar{\bf x},\hat{\bf n})^2 = \sum_{i=1}^{3}\left(c_i(\bar{\bf x})n_i\right)^2\, .  
\label{eq:elliptical}
\end{equation}
Therefore, setting eqs.~(\ref{eq:elliptical}) and~(\ref{eq:dvdn_grupo}) in eq.~(\ref{gruprel2}), we have
\begin{equation}
\cos\psi(\hat{\bf n}) = \frac{\sum_{i=1}^3 (c_i(\bar{\bf x})n_i)^2}{\sqrt{\sum_{i=1}^3 (c_i^2(\bar{\bf x})n_i)^2}}\, .   
\end{equation}
\cite{Tsvankin:2001} comments that an elliptical-slowness surface leads to an ellipsoidal wavefront from a point source (ray-velocity surface) and this property is intrinsic to SH-wave anisotropy in transversely isotropic media. 

In conclusion, the relationship given by eq.~(\ref{gruprel2}) is greatly advantageous to our purposes, which consists of studying high-order approximations of the wavefront in the vicinity of the central ray and understanding how the geometrical spreading of this wavefront actually works. It is worth to mention that, for this approach, the use of Cartesian coordinates is not adequate to show some particularities of physical-mathematical properties. Thus, in the following sections, we use one of the approaches through ray-centered coordinates, which we call the ray-centered physical coordinates, and derive an explicit expression for the Hamiltonian and its applications in terms of these coordinates in a reduced system. Moreover, explicit velocity and displacement-slowness expressions allow us to study wave phenomena in the context of specific materials. In particular, we use these expressions to formulate inverse problems where elasticity parameters are calculated based on slope and curvature information from traveltimes obtained from experimental measurements.

\section{Formulation in ray-centered coordinates}
\label{sec:3}

Dynamic ray tracing in centered coordinates uses a vectorial base updated by ordinary differential equations to obtain wavefront information along the raypath. This ray-centered base yields the first derivatives of phase-space coordinates of a ray point concerning initial conditions. Second- or higher-order spatial traveltime derivatives can also be computed and are essential to calculate ray perturbations concerning initial conditions or parameter-model variations, mainly for two-point ray tracing, and the paraxial-ray approximation approaches \citep{Klime1994}. To make this approach more accessible, we present Lagrangian and Hamiltonian formulations of the paraxial-ray theory in a non-necessarily orthogonal ray-centered coordinates system, assuming that the ray trajectories across any smooth heterogeneous and anisotropic medium. 

We start by assuming a raypath described by the system of differential equations, eq.~(\ref{eq:systemEDO}), and with the position-vector curve parameterized by the arc length of the ray, represented here by the letter $s$, it is defined by
\begin{equation}
 \bar{\mathbf{x}}(s)=x_1(s)\hat{\mathbf{i}}_1 + x_2(s)\hat{\mathbf{i}}_2 + x_3(s)\hat{\mathbf{i}}_3\, ,
\label{centralray}
\end{equation}
where $\hat{\bf i}_1$, $\hat{\bf i}_2$, and $\hat{\bf i}_3$ are rigid versors that define a 3D Cartesian coordinate system. Therefore, assuming that $\bar{\bf x}(s)$ is a smooth curve parameterized by its arc length. 
Thus, by differential geometry, for any point on the raypath, it is possible to define a unit vector function as
\begin{equation}
\hat{\mathbf{t}}(s)=\frac{d \bar{\mathbf{x}}}{ds}(s).
\label{tang00}
\end{equation}
Then, by definition, at any point on the raypath ${\bar{\bf x}}(s)$, there is a unit vector $\hat{\bf t}$ that is tangent to the curve at this point.
Besides, from eqs.~(\ref{groupvel}) and~(\ref{tang00}), and on the ray-velocity vector, we take the following derivation
\begin{equation}
\frac{d\bar{\bf x}}{dt}=\frac{d\bar{\bf x}}{ds}\frac{ds}{dt}=v_g\hat{\bf t}\, ,
\label{eq:vg_t}
\end{equation}
concerning the physical meaning of these elements, we observe that $\hat{\bf t}$ indicates the normalized group velocity direction movement while the scalar, 
\begin{equation}
v_g = \frac{ds}{dt}\, ,
\label{eq:gvelocity}
\end{equation}
is the group velocity module, as the ratio between the infinitesimal propagation along the arclength by time.

In order to introduce the paraxial-ray coordinates, we define ray-centered coordinates $\bar{\bf q} = [q_1,q_2,s]^T$ along a particular ray. As already mentioned, we parametrize the points along the ray by the (arc length) monotonic variable $s$. Again, at each point $\bar{\bf x}$, on curve $\bar{\bf x}(s)$ that describes the ray trajectory, we choose two orthonormal vectors, $\hat{\bf e}_1$ and $\hat{\bf e}_2$, perpendicular to phase unit vector $\hat{\bf n}$ at that point. We denote the central ray as the trajectory of this principal ray and the trajectories in its vicinity of paraxial rays. Therefore, the ray-centered coordinate system consists of a curvilinear coordinate $\bar{\bf q}$ related to the central ray, on which $q_1 = q_2 = 0$, and the third coordinate changes monotonically on the central ray.

\begin{figure*}
\centering
\includegraphics[width=0.65\textwidth]{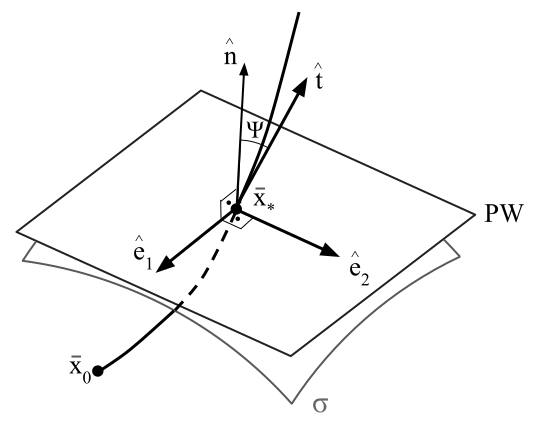}
\caption{Ray-centered coordinates $\{\hat{\mathbf{e}}_{1}(s),\hat{\mathbf{e}}_{2}(s),\hat{\mathbf{t}}(s)\}$ at a given point along the ray related to the wavefront $\sigma$ tangent to the plane wave $PW$. The angle $\psi$ represents the anisotropy deviation in the wave propagation phenomena.}
\label{fig:2}
\end{figure*}

Based on the aforementioned parametric construction, we introduce a parametrized unitary system with a centered base $\{\hat{\mathbf{e}}_{1}(s),\hat{\mathbf{e}}_{2}(s),\hat{\mathbf{t}}(s)\}$, where $\hat{\bf e}_{1}(s)$ and $\hat{\bf e}_{2}(s)$ obeying the relations, respectively,
\begin{equation}
 \frac{d\hat{\mathbf{e}}_{1}}{ds}(s)=h_{1}(s)\hat{\mathbf{n}}(s) \quad \text{and} \quad \frac{d\hat{\bf e}_{2}}{ds}(s)=h_{2}(s)\hat{\bf n}(s)\, .
 \label{deriv}
\end{equation}
We describe later the physical quantities of the vector inputs ${\bf h} = [h_1(s),h_2(s)]^T$, more detail, see in section~\ref{sec:h}. Through the ray-centered coordinate system, $\bar{\bf q}$, we define a position vector in the paraxial ray determined by $\bar{\bf r}(\bar{\bf q})$ and written as \citep{Popov:1978,Klime1994}
\begin{equation}
 \bar{\bf r}(\bar{\bf q})=\bar{\bf x}(s)+q_{1}\hat{\bf e}_{1}(s)+q_{2}\hat{\bf e}_{2}(s)\, ,
\label{paraxialpoint}
\end{equation}
i.e., the transformation from the ray–centered coordinates $\bar{\bf q}$ to Cartesian coordinates $\bar{\bf r}$. Figure \ref{fig:2} illustrates the coordinate system described above.
We can observe that the first two vectors $\hat{\bf e}_{1}$ and $\hat{\bf e}_{2}$ of the ray-centered coordinate system define a plane-wave, $PW$, tangent to the wavefront, $\sigma$, at $\bar{\bf x}$, and the third vector $\hat{\bf t}$ is tangent to the central ray at some point. Indeed, there are other ray-centered basis configurations, but we chose the basis $\{\hat{\bf e}_{1}(s), \hat{\bf e}_{2}(s), {\hat{\bf t}(s)}\}$ to explore its physical properties. Indeed, this base construction carries information about the wavefront plane and its raypath direction on a central ray. Furthermore, throughout  this work, we denote this physical formulation of ray-centered coordinates as the ray-centered coordinate system. 

Now, to compute the differential of paraxial-ray position straightforwardly, we take the differential of the position vector at paraxial-ray coordinates, which yields
\begin{equation}
d\bar{\bf r} = \left(\hat{\bf t} + \left({\bf h}\cdot{\bf q}\right)\hat{\bf n}\right)ds + \hat{\bf e}_1 dq_1 + \hat{\bf e}_2 dq_2\, , 
\label{vicinitypoint}
\end{equation}
where the reduced ray-centered coordinates ${\bf q}=[q_1,q_2]^T$ determines the plane wavefront coordinates. From that, we define the area element $(d\sigma)^2$, valid in the vicinity of the central ray, in relation to the differential $d\bar{\bf r}$ as
\begin{equation}
(d\sigma)^2 =  d\bar{\bf r}\cdot d\bar{\bf r}.
\label{varaniso}
\end{equation}
%

Besides, by construction, we have the following relationships between coordinates vectors
\begin{equation}
\hat{\bf t}\cdot\hat{\bf n} = \cos\psi\, , \quad
\hat{\bf t}\cdot\hat{\bf e}_1 = \sin\psi\cos\phi\, , \quad
\hat{\bf t}\cdot\hat{\bf e}_2 = \sin\psi\sin\phi\, . 
\label{eq:pvetores}
\end{equation}
It implies that the differential of a paraxial-ray position of wavefront propagation can be described as
\begin{equation}
d\sigma = \sqrt{\chi^2 ds^2 + 2\left(d{\bf q}\cdot{\bf a}_\phi\right)\sin\psi ds + (d{\bf q}\cdot d{\bf q})^2}\, ,
\label{area}
\end{equation}
with
\begin{equation}
\chi^2 = \chi(\bar{\bf q})^2 = 1 + 2\left({\bf h}\cdot{\bf q}\right)\cos\psi + \left({\bf h}\cdot{\bf q}\right)^2\, , 
\label{areadif}
\end{equation}
where ${\bf a}_\phi=[\cos\phi,\sin\phi]^T$ is the ray-centered azimuthal vector. Also, the ray-centered coordinates have a specific region of validity in the vicinity of the central ray, arising from the fact that for a curved ray, different values for $\bar{\bf q}$ can result in the same value for $\bar{\bf r}$. Hence, there is no one-to-one correspondence between ray-centered and Cartesian coordinates for greater paraxial distances. Thus, to discover this region of validity, we use eq.~(\ref{area}) and observe that the components of metrical tensor $g_{ij}$ of the ray-centered coordinate system have the following relations
\begin{equation}
\left\{
\begin{split}
g_{11} = g_{22} &= 1\, ,\\
g_{12} = g_{21} &= 0\, ,\\
g_{13} = g_{31} &= \sin\psi\cos\phi\, , \\ 
g_{23} = g_{32} &= \sin\psi\sin\phi\, , \\
g_{33} &= \chi^2\, .
\end{split} \right.
\label{eq:gij}
\end{equation}
We can use variational principles \citep{Lanczos:1986} to take an alternative approach to the definition of rays. However, for the raypaths to be geodesic in a Riemannian space with the metrical tensor given by eq.~(\ref{eq:gij}) is sufficient that
\begin{equation}
\cos\psi > |{\bf h}\cdot{\bf q}|\, .   
\label{eq:invariant}
\end{equation}
In other words, if the condition in eq.~(\ref{eq:invariant}) is satisfied, then the kinematic properties of ray tracing are invariant under the transformation of coordinates given by eq.~(\ref{paraxialpoint}). Therefore, we have that the differential $d\sigma$ gives the distance between two adjacent wavefronts, and $ds$ measures the two-point raypath distance as shown in Figure~\ref{fig:1}.

From eqs.~(\ref{gruprel2}) and~(\ref{eq:gvelocity}), we describe the two-point raypath traveltime by the following functional action
\begin{equation}
t = \int_{0}^{s_*} \frac{ds}{v_{g}} =  \int_{0}^{s_*} \frac{(\cos\psi) ds}{v}\, .
\label{fermat_aniso}
\end{equation}
For a given two-point raypath of length $s_*$, the first equality of eq.~(\ref{fermat_aniso}) describes the traveltime of the propagation phenomenon from the standpoint of the group velocity on this raypath. The second equality concerns the traveltime variation as a function of the phase velocity and the weight function given by cosine between the group and phase vectors. 

Given a central ray connecting two points, $\bar{\bf x}_0$ and $\bar{\bf x}_*$, we show that this raypath is a stationary traveltime trajectory on any other paraxial ray from $\bar{\bf x}_0$ to $\bar{\bf r}_*$ over the plane, $PW$, generated by the unit vectors $\hat{\bf e}_1(s_*)$ and $\hat{\bf e}_2(s_*)$ on $\bar{\bf x}_*$, as shown in Figure~\ref{fig:2}. Therefore, based on the observations mentioned earlier, we first analyze under what conditions the functional action integrals obey
\begin{equation}
\int_{0}^{s_*}\frac{(\cos\psi)ds}{v} = \int_{0}^{s_*} \frac{d\sigma}{ds}\frac{ds}{v} \, .
\label{perturbvel}
\end{equation}
Nonetheless, for such equality to be true to any value $v$, we start by taking $d\sigma/ds=\cos\psi$. In mathematical terms, and taking $\dot{\bf q} = d{\bf q}/ds$ to simplify the notation, we have
\begin{equation}
\cos^{2}\psi = \chi^2 + 2\left({\bf a}_\phi\cdot\dot{\bf q}\right)\sin\psi + \left(\dot{\bf q}\cdot\dot{\bf q}\right)^2\, .
\end{equation}
On the central ray, i.e., ${\bf q} = {\bf 0}$, which results $\chi=1$ in the previous expression and yields 
\begin{equation}
\cos^{2}\psi= 1 + 2\left({\bf a}_\phi\cdot\dot{\bf q}\right)\sin\psi + \left(\dot{\bf q}\cdot\dot{\bf q}\right)^2\, ,
\end{equation}
which is, in turn, valid if and only if
\begin{equation}
\left(\dot{\bf q}+\sin\psi{\bf a}_\phi\right)\cdot\left(\dot{\bf q}+\sin\psi{\bf a}_\phi\right)=0\, ,
\end{equation}
which ultimately implies
\begin{equation}
\left.\dot{\bf q}\right|_{{\bf q}={\bf 0}}= \dot{\bf q}_0 = -{\bf a}_\phi\sin\psi\, .
\label{dqds}
\end{equation}
When evaluated on the central ray, we can observe that quantity $|\sin\psi|$ quantifies the anisotropic deviation influence on $\dot{\bf q}$.

Therefore, from eqs.~(\ref{fermat_aniso}) and~(\ref{dqds}), in the central-ray direction, i.e., $({\bf q},\dot{\bf q})=({\bf 0},\dot{\bf q}_0)$, the following equality is obeyed
\begin{equation}
\int_{0}^{s_*} \frac{ds}{v_{g}} = \int_{0}^{s_*}\frac{d\sigma}{ds}\frac{ds}{v}\, ,
\label{fermat_aniso2}
\end{equation}
and with that, we guarantee the uniqueness of this raypath from two-point, from $\bar{\bf x}_0$ to $\bar{\bf x}_*$, with the same departure and arrival raypath-tangent vectors.

In order to guarantee that the raypath in the central-ray direction obeys the Fermat principle, in other words, the stationary-action principle, we set the traveltime action functional as
\begin{equation}
t({\bf q}) = \int_{0}^{s_*} \left(\sqrt{\chi^2  + 2\sin\psi\left(\dot{\bf q}\cdot{\bf a}_\phi\right) +\dot{\bf q}\cdot \dot{\bf q}}\right) \frac{ds}{v}\, ,
\label{eq:t_q}
\end{equation}
where the Lagrangian ${\cal L}_s$ is defined as
\begin{equation}
{\cal L}_s\left({\bf q},\dot{\bf q}\right) = \frac{1}{v}\frac{d\sigma}{ds} = \frac{1}{v}\sqrt{\chi^2  + 2\sin\psi\left(\dot{\bf q}\cdot{\bf a}_\phi\right) +\dot{\bf q}\cdot \dot{\bf q}}\, .
\label{lagrange}
\end{equation}
Therefore, to ensure that the solution is stationary in $t$, we need to demonstrate that ${\cal L}_s$ obeys the Euler-Lagrange equation in the ray-central direction. To help us in this proof, we introduce the conjugate-type momenta ${\bf p} = [p_1, p_2]^T$, by definition, can be written as
\begin{equation}
 {\bf p}=\frac{\partial {\cal L}_s}{\partial \dot{\bf q}}\, .
 \label{lagrange-slownessa}
\end{equation}
Therefore, by the Euler-Lagrange equation, for $t({\bf 0})$ to be a stationary solution to eq.~(\ref{eq:t_q}), such relation must occur \citep{Gelfand:2000}
\begin{equation}
\frac{\partial {\cal L}_s}{\partial {\bf q}} - \frac{d}{ds}\left(\frac{\partial {\cal L}_s}{\partial \dot{\bf q}}\right) = {\bf 0}\, ,
\label{eq:EL}
\end{equation}
in the central-ray direction. From eq.~(\ref{lagrange-slownessa}), we have the following value of ${\bf p}$ in the central-ray direction as
\begin{equation}
{\bf p}=\left.\frac{\partial{\cal L}_s}{\partial \dot{\bf q}}\right|_{({\bf 0},\dot{\bf q}_0)} = {\bf 0}\, ,    
\end{equation}
where for eq.~(\ref{eq:EL}) results to be true, it is necessary that
\begin{equation}
\frac{d{\bf p}}{ds}=\left.\frac{\partial{\cal L}_s}{\partial {\bf q}}\right|_{({\bf 0},\dot{\bf q}_0)} = {\bf 0}\, .  
\label{dpds}
\end{equation}

By construction, see eq.~(\ref{eq:pvetores}), $\partial\psi/\partial{\bf q}$ and $\partial{\bf a}_\phi^T/\partial{\bf q}$ are null vector and matrix, respectively, for any value of ${\bf q}$. Also, from eq.~(\ref{paraxialpoint}), on the central ray, we can write
\begin{equation}
\frac{\partial v}{\partial q_i} = \left.\frac{\partial\bar{\bf r}}{\partial q_i}\cdot\frac{\partial v}{\partial\bar{\bf r}}\right|_{{\bf q}={\bf 0}} = \hat{\bf e}_i\cdot\frac{\partial v}{\partial\bar{\bf x}}\, , \quad i=1,2\, .
\label{eq:dvdx}
\end{equation}
Therefore, in the central-ray direction condition
\begin{equation}
\left.\frac{\partial{\cal L}_s}{\partial {\bf q}}\right|_{({\bf 0},\dot{\bf q}_0)} = \frac{1}{v_0}\left({\bf h}_0 - \frac{\cos\psi_0}{v_0}\frac{\partial v_0}{\partial {\bf q}}\right)\, , 
\end{equation}
where $v_0$, $\cos\psi_0$, ${\bf h}_0$, and $\partial v_0/\partial{\bf q}$ are evaluated in the central-ray direction.

Then, to have condition in eq.~(\ref{dpds}) satisfied it is necessary that 
\begin{equation}
{\bf h}_0(s) = \frac{\cos\psi_0}{v_0}\frac{\partial v_0}{\partial {\bf q}}\, . 
\label{curv}
\end{equation}

In order to prove the result of eq.~(\ref{curv}), it is necessary to pay attention that over any central ray, by definition, we have $\bar{\bf w}\cdot\hat{\bf e}_i = 0$. Therefore, differentiating with respect to the time variable gives
\begin{equation}
\frac{d}{dt}\left(\bar{\bf w}\cdot\hat{\bf e}_i\right) = \frac{d\bar{\bf w}}{dt}\cdot\hat{\bf e}_i + \bar{\bf w}\cdot\frac{d\hat{\bf e}_i}{dt} = 0 
\end{equation}
and changing it with relation eqs.~(\ref{eq:systemEDO}), (\ref{eq:gvelocity}), and~(\ref{deriv}), we obtain 
\begin{equation}
\frac{{\bf h}}{\cos\psi} = \left(\frac{1}{v}\frac{\partial v}{\partial {\bf q}}\right)_{{\bf q}={\bf 0}}\, , 
\label{eq:hcos}
\end{equation}
for any $\dot{\bf q}$-value on a central ray. Finally, in the central-ray direction, eq.~(\ref{eq:hcos}) proves the necessary statement.

On these physical trajectories with length, $ds$ and $d\sigma$, the ray traveltime and wavefront distance are optimized simultaneously, as seen in Figure~\ref{fig:1}. Moreover, as done in eq.~(\ref{tang00}), on the central ray, we have from eq.~(\ref{varaniso}) that
\begin{equation}
\frac{d\bar{\bf r}}{d\sigma} = \frac{d\bar{\bf r}}{ds}\left(\frac{d\sigma}{ds}\right)^{-1} = \hat{\bf n}    
\end{equation}
gives the unit vector orthogonal to the plane-wave tangent to the wavefront. Also, we have the minimal distance of a paraxial ray from point $\bar{\bf x}_0$ to the plane wavefront containing point $\bar{\bf x}_*$ is the differential $d\sigma = d\bar{\bf r}_*$.

\subsection{Hamiltonian formulation}

A simple interpretation of Hamiltonian mechanics in seismic wavefront applications comes from its interpretation in paraxial-ray theory, which describes the dynamic properties of traveltime trajectories \citep{Cerveny:2001}. By Legendre transformation \citep{Gelfand:2000}, the Hamiltonian expression in its reduced centered coordinates $({\bf q}, {\bf p})$ is given by (see, Appendix~\ref{Appendix:Hamiltonian})
\begin{equation}
{\cal H}_s({\bf q}, {\bf p}) = -\frac{1}{v}(\cos\psi + {\bf h} \cdot\mathbf{q})\sqrt{1-v^2({\bf p}\cdot{\bf p})}-\sin\psi({\bf p}\cdot{\bf a}_\phi)\, . 
\label{hamilton}
\end{equation}
Therefore, applying the Legendre transformation to the Lagrangian function ${\cal L}_s$, considering $\dot{\bf q}$ as active variables of the transformation and the position coordinate $\bar{\bf q}$ as passive variable. The $\dot{\bf q}$ value is transformed into the conjugate-type momenta ${\bf p}$, and the Lagrangian function is transformed into the Hamiltonian function ${\cal H}_s$, which runs over the arc-length parameter. Therefore, its Hamiltonian system is given by 
\begin{equation}
 \frac{d{\bf q}}{ds}=\frac{\partial {\cal H}_s}{\partial {\bf p}}\, , \quad \text{and} \quad \frac{d{\bf p}}{ds}=-\frac{\partial {\cal H}_s}{\partial {\bf q}}\, ,
\label{sisthamilton}
\end{equation}
where
\begin{equation}
\begin{split}
\frac{\partial {\cal H}_s}{\partial {\bf p}} &= \frac{v\left(\cos \psi +{\bf h}\cdot \mathbf{q}\right){\bf p}}{\sqrt{1-v^{2}({\bf p}\cdot{\bf p})}}-\sin\psi{\bf a}_\phi\, ,\\ 
\frac{\partial {\cal H}_s}{\partial {\bf q}} &= \frac{1}{v^2}\frac{\partial v}{\partial{\bf q}}\frac{\left(\cos \psi +{\bf h}\cdot \mathbf{q}\right)}{\sqrt{1-v^{2}({\bf p}\cdot{\bf p})}} - \frac{{\bf h}}{v}\sqrt{1-v^{2}({\bf p}\cdot{\bf p})}\, .
\end{split}
\label{sisthamilton2}
\end{equation}

The main reason we work with Hamiltonian equations instead of Lagrangian equations is the property of the number of variables being doubled. This increase in the number of variables makes it possible to expand the field of possible transformations of coordinates, having a more significant number of variables at our disposal. However, we do not possess any systematic method for simplifying the Lagrangian function in Lagrangian mechanics. Meanwhile, coordinates can be transformed in Hamiltonian mechanics to systematically produce ignorable variables and simplify the Hamiltonian function. Therefore, with the help of this reduction procedure, we change an integration problem with six distinct equations, $(\bar{\bf r}(s),d\bar{\bf r}/ds)$, into just four. Furthermore, the system represented in four-dimensional space by the curve $({\bf q}(s),{\bf p}(s))$ are the solutions to the problem on the paraxial ray. Thus, following the conditions established in eqs.~(\ref{dqds}) and~(\ref{dpds}), we have the Hamiltonian in the central-ray direction as
\begin{equation}
\left.\frac{\partial {\cal H}_s}{\partial {\bf p}}\right|_{({\bf 0},{\bf 0})} = -\sin\psi_0{\bf a}_\phi\, , \quad \text{and} \quad \left.\frac{\partial {\cal H}_s}{\partial {\bf q}}\right|_{({\bf 0},{\bf 0})} = {\bf 0}\, .
\label{eq:sisthamilton0}
\end{equation}

\subsection{Raypath curvature}
\label{sec:h}

In general, ray paths are curves. In isotropic media, the raypath curvature is known to be \citep{Popov:2002}
\begin{equation}
|\kappa(s)| = \frac{1}{v_0}\left\|\frac{\partial v_0}{\partial{\bf q}}\right\|\, , 
\end{equation}
where $\kappa(s)$ is the raypath curvature at $\bar{\bf x}(s)$. Therefore, we determined that quantity from the paraxial-ray tracing system in the ray-centered coordinate for general anisotropic media.

\begin{figure*}
\centering
\includegraphics[width=0.65\textwidth]{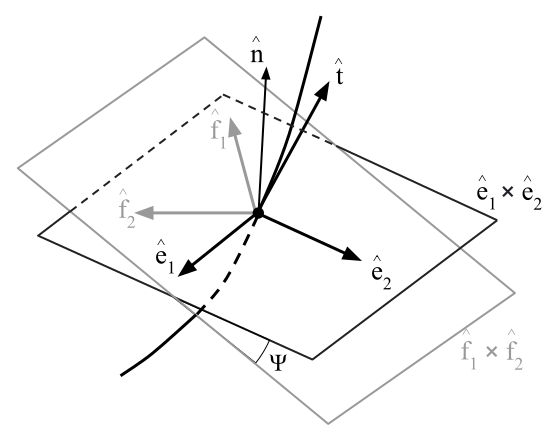}
\caption{Normal planes related to the phase and group vectors, $\hat{\bf n}$ and $\hat{\bf t}$, respectively. Note that both planes differ from an angle $\psi$.}
\label{fig:77}
\end{figure*}
At the initial point in the central-ray direction, the direction of vector $\hat{\bf e}_1$ can, in principle, be chosen arbitrarily in the normal plane to the slowness vector. Based on this premise, we consider the start unit vector $\hat{\bf e}_1(0)$ as any vector orthogonal to $\hat{\bf n}(0)$, and we construct the second-orthogonal vector to both, using the cross product $\times$, as
\begin{equation}
\hat{\bf e}_2 = \hat{\bf n}\times\hat{\bf e}_1\, .   
\end{equation}
Thus, we constructed the standard option for the orthonormal wavefront bases of the ray-centered coordinate system, the vectors $\hat{\bf e}_1$ and $\hat{\bf e}_2$. Furthermore, we take two orthogonal vectors that generate the normal plane to the group vector $\hat{\bf t}$ as
\begin{equation}
\bar{\bf f}_1 = \hat{\bf t}\times\hat{\bf e}_2\, , \quad \bar{\bf f}_2 = \hat{\bf e}_1\times\hat{\bf t}\, ,  
\label{eq:f1f2}
\end{equation}
which arrives in (see, Figure \ref{fig:77})
\begin{equation}
\hat{\bf e}_i\cdot\bar{\bf f}_i = -\cos\psi\, , \quad
\hat{\bf n}\cdot\bar{\bf f}_1 = \sin\psi\cos\phi\, , \quad
\hat{\bf n}\cdot\bar{\bf f}_2 = \sin\psi\sin\phi\, .
\label{trasnf}
\end{equation}
We can observe that if the anisotropic deviation is null, i.e., $\psi = 0$, then the vectors $\hat{\bf e}_{1}$, $\hat{\bf e}_{2}$, $\bar{\bf f}_{1}$, and $\bar{\bf f}_{2}$ belong to the same normal plane to the slowness vector. 

Now, differentiating the unit vector of the ray, $\hat{\bf t}$, with respect to $s$, we get the raypath curvature multiplied by some unit vector on the plane generated by $\hat{\bf f}_1$ and $\hat{\bf f}_2$, namely
\begin{equation}
\frac{d\hat{\bf t}}{ds} = \kappa\hat{\bf f}_{12} = \kappa\left((\hat{\bf f}_1\cdot\hat{\bf f}_{12})\hat{\bf f}_1 + (\hat{\bf f}_2\cdot\hat{\bf f}_{12})\hat{\bf f}_2\right)\, ,   
\end{equation}
where $\kappa(s)$ is the curvature of ray at $\bar{\bf x}(s)$ and $\hat{\bf f}_{12}$ is the unit vector orthogonal to $\hat{\bf t}$ such as 
\begin{equation}
(\hat{\bf f}_1\cdot\hat{\bf f}_{12})^2 + (\hat{\bf f}_2\cdot\hat{\bf f}_{12})^2 = 1\, .    
\end{equation}

To obtain the raypath curvature value, we set 
\begin{equation}
[\hat{\bf e}_1; \hat{\bf e}_2]^T\hat{\bf t} = -\dot{\bf q}_{0}(s) 
\end{equation}
and by Appendix~\ref{appendix:H2}, in the central-ray direction, we have
\begin{equation}
\frac{\partial}{\partial{\bf p}}\left(\frac{d{\cal H}_s}{ds}\right) = \frac{d}{ds}\left(\frac{\partial{\cal H}_s}{\partial{\bf p}}\right) = \frac{d\dot{\bf q}_0}{ds} = \ddot{\bf q}_0(s) = {\bf 0}\, .  
\label{eq:qdd}
\end{equation}
Therefore, in the central-ray direction, we arrived at
\begin{equation}
\begin{split}
\frac{d}{ds}\left(\hat{\bf t}\cdot\hat{\bf e}_i\right) &= \hat{\bf t}\cdot\frac{d\hat{\bf e}_i}{ds} + \hat{\bf e}_i\cdot\frac{d\hat{\bf t}}{ds} \\
&= \cos\psi_0\left(h_i - (\hat{\bf f}_i\cdot\hat{\bf f}_{12})\frac{\kappa}{\|\bar{\bf f}_i\|}\right) = 0\, , \ i=1,2,
\end{split}
\end{equation}
i.e., the curvature of the ray depends upon velocity and its derivatives as follows
\begin{equation}
|\kappa| = \sqrt{\sum_{i=1}^2\left(\|\bar{\bf f}_i\|h_i\right)^2}\, .
\label{eq:kappa}
\end{equation}
As $\cos\psi_0$ and $v_0$ are always non-null numbers, eq.~(\ref{eq:kappa}) implies that the following theorem,
\begin{equation}
\kappa = 0 \iff \ \frac{\partial v_0}{\partial {\bf q}} = {\bf 0}\, ,  
\end{equation}
is valid for any heterogeneous anisotropic medium. In other words, the variation of the slowness vector $d\bar{\bf w}/dt$ is orthogonal to the wavefront, if only if the raypath curvature is zero. Furthermore, from eq.~(\ref{eq:invariant}), the raypath curvature influences the width of the region of validity that makes the coordinate transformation one-to-one.

\section{Dynamic Ray Tracing system}

Various coordinate systems can represent dynamic ray-tracing systems for a general anisotropic medium. The most convenient and frequently adopted is the wavefront-orthonormal coordinate system \citep{Klime1994, Cerveny:2001} and the Cartesian coordinate system \citep{Cerveny:1972, Iversen:2021}. In this section, we use the Hamiltonian system, defined in the previous sections, to analyze the dynamic properties along the central ray using a non-orthogonal system. Unlike the Cartesian coordinate system, which consists of six ordinary linear equations, the ray-centered formulation can reduce to only four linear equations \citep{Bliss:1916}.

The approach starts with Cartesian coordinates and, via coordinate transformation, generates a reduced coordinate system that solves the mathematical difficulties associated with numerical modeling. However, describing the derivations directly from the Hamiltonian in a physical system of centered coordinates explains, more clearly, the physical properties of the dynamical problem. Furthermore, the wavefront-orthonormal coordinate system can describe the system with mathematical precision. Although without any explicit physical information.
On the other side, our approach allows an understanding of the physical phenomenon since the anisotropic correction factor is explicit in the formulation. Therefore, analyzing phase velocity variation concerning direction and position allow us to fully characterize the contributions of anisotropy and heterogeneity in the paraxial-ray formulation.
 
\subsection{Geometrical spreading}

Historically, the concept of geometrical spreading plays a crucial role in the computation of amplitudes related to seismic body waves. Commonly, geometrical spreading is introduced concerning the cross-sectional area of the ray tube or in some relation to the ray Jacobian \citep{Popov:2002}. Unfortunately, the definition of geometrical spreading in the seismological literature is not unique. Concerning our studies, we follow the definition given by \cite{Cerveny:2001}.

Geometrical spreading is the phenomenon of energy scattering over a surface due to the expansion or contraction of its wavefronts. Such a geometrical deformation is independent of frequency and significantly affects almost all situations of propagating ray vectors. In order to understand these problems and to determine that geometrical deformation, we remark that the set of rays and wavefronts form an orthogonal curvilinear coordinate system. Moreover, for a set of rays and wavefronts from a curvilinear coordinate system, we can make the following parameterization $\bar{\bf r} = \bar{\bf r}(\bar{\bf q}(\bar{\boldsymbol{\gamma}}))$ where $\bar{\boldsymbol{\gamma}} = [\gamma_1,\gamma_2,t]^T$ and with
\begin{equation}
\frac{\partial\bar{\bf q}}{\partial\gamma_i}\cdot\frac{\partial\bar{\bf q}}{\partial t} = 0\, , \quad i=1,2. 
\label{eq:dqdt_dqdgamma}
\end{equation}
So, by \cite{Cerveny:2001}, we can mathematically define the geometrical spreading through the Jacobian ${\cal J}$ as
\begin{equation}
\Lambda(\bar{\boldsymbol{\gamma}}) = \sqrt{\frac{1}{v_g}|{\cal J}(\bar{\boldsymbol{\gamma}})|} = \left|\frac{dt}{ds}\det\left[\frac{\partial\bar{\bf r}^T}{\partial\bar{\boldsymbol{\gamma}}}\right]\right|^{\frac{1}{2}}\, . \label{eq:GS}  
\end{equation}
Therefore, each parameter $\boldsymbol{\gamma} = [\gamma_1,\gamma_2]^T$ defines a ray and the $t$-value is the running parameter. The curvilinear coordinate set $\bar{\boldsymbol{\gamma}}$ defined in this form is usually called the local ray coordinate system. Any point in the region illuminated by rays may be defined by its ray coordinates. In other terms, ${\cal J}$ is the Jacobian of the transformation from Cartesian to local ray coordinates. Finally, in any position of a ray tube volume ${\cal V}$, we can measure its volume element in this coordinate system as
\begin{equation}
d{\cal V} = {\cal J}(\bar{\boldsymbol{\gamma}}) v_g dt d\gamma_1 d\gamma_2 \, .     
\end{equation}

To compute the geometrical spreading, $\Lambda$, let us consider a system of rays, parameterized by a ray parameter $\boldsymbol{\gamma}$ such that ${\bf q}={\bf q}(\boldsymbol{\gamma})$ and ${\bf p}={\bf p}(\boldsymbol{\gamma})$ are solutions starting from a point $\boldsymbol{\gamma}$ on an initial neighborhood representing the ray tube. In order to make such calculations, we take the derivatives of the Hamiltonian system with respect to these parameters as
\begin{equation}
\begin{split}
\frac{d}{d t}\left[ \frac{\partial {\bf q}^T}{\partial \boldsymbol{\gamma}}\right] &= \frac{\partial }{\partial \boldsymbol{\gamma}}\left[\frac{ds}{dt}\dot{\bf q}\right]^T =\frac{\partial }{\partial\boldsymbol{\gamma}}\left[\frac{v}{\cos\psi}\frac{\partial{\cal H}_s}{\partial {\bf p}}\right]^T\, , \\
\frac{d}{d t}\left[ \frac{\partial{\bf p}^T}{\partial\boldsymbol{\gamma}}\right] &=  \frac{\partial }{\partial\boldsymbol{\gamma}}\left[\frac{ds}{d t}\dot{\bf p}\right]^T=-\frac{\partial }{\partial\boldsymbol{\gamma}}\left[\frac{v}{\cos\psi}\frac{\partial{\cal H}_s}{\partial \bf{q}}\right]^T\, .
\end{split}
\end{equation}
Therefore, taking the following vectors as
\begin{equation}
{\bf m}_p = \frac{v}{\cos\psi}\frac{\partial{\cal H}_s}{\partial{\bf p}} = \frac{d{\bf q}}{dt}\, , \quad {\bf m}_q = \frac{v}{\cos\psi}\frac{\partial{\cal H}_s}{\partial{\bf q}} = -\frac{d{\bf p}}{dt}\, ,
\label{eq:mqmp}
\end{equation}
where $v$, $\psi$, and ${\bf a}_\phi$, also, are functions of ${\bf p}$. Also, denoting ${\bf Q}=\partial {\bf q}^T/\partial\boldsymbol{\gamma}$ and ${\bf P}=\partial{\bf p}^T/\partial\boldsymbol{\gamma}$, and using the chain rule, the above system can be written in a matrix form, which represents the propagator matrix in the central-ray direction \citep{Cerveny:2001}, given by
\begin{equation}
\frac{d}{dt} \left[\begin{matrix}
\mathbf{Q}\\
\mathbf{ P}
\end{matrix}\right]=\left[\begin{matrix}
\frac{\partial {\bf m}_p^T}{\partial{\bf q}}& \frac{\partial {\bf m}_p^T}{\partial{\bf p}} \\
-\frac{\partial {\bf m}_q^T}{\partial{\bf q}} & -\frac{\partial {\bf m}_q^T}{\partial{\bf p}}
\end{matrix}\right]  \left[\begin{matrix}
\mathbf{Q}\\
\mathbf{P}
\end{matrix}\right]\, .
\label{seconsystem}
\end{equation}  

Finally, to compute the values of the third coordinate derivatives, from eqs.~(\ref{eq:vg_t}), (\ref{paraxialpoint}), and~(\ref{eq:EF}) in central ray direction, we take 
\begin{equation}
\left.\frac{\partial\bar{\bf q}}{\partial t}\right|_{\bar{\boldsymbol{\gamma}}_0} = \frac{\partial s}{\partial t}\left(\frac{\partial\bar{\bf r}^T}{\partial\bar{\bf q}}\right)^{-T}\frac{\partial\bar{\bf r}}{\partial s} = v_g\bar{\bf F}_0\hat{\bf t} = [0, 0, v_g]^T\, ,    
\end{equation} 
Therefore, from the property given in eq.~(\ref{eq:dqdt_dqdgamma}), yields $\partial s/\partial\bar{\boldsymbol{\gamma}} = [0,0,v_g]^T$, which implies in 
\begin{equation}
\left.\frac{\partial \bar{\bf r}^T}{\partial\bar{\boldsymbol{\gamma}}}\right|_{\bar{\boldsymbol{\gamma}}_0} = \bar{\bf E}_0\left.\frac{\partial \bar{\bf q}^T}{\partial\bar{\boldsymbol{\gamma}}}\right|_{\bar{\boldsymbol{\gamma}}_0} = \bar{\bf E}_0\left[\begin{matrix}
{\bf Q}& {\bf 0}^T \\
{\bf 0} & v_g
\end{matrix}\right]\, ,
\end{equation}
and from definition of geometrical spreading given by eq.~(\ref{eq:GS}), we obtain
\begin{equation}
\Lambda(\bar{\boldsymbol{\gamma}}_0) = \sqrt{\cos\psi|\det{\bf Q}|}\, . 
\label{eq:geospreading}
\end{equation} 

\subsection{Propagation matrix description}

In order to determine well-defined expressions for the elements of the propagator matrix in the central-ray direction is necessary to compute the propagation matrix showed in eq.~(\ref{seconsystem}) using the definition given by eq.~(\ref{eq:mqmp}). However, it is necessary to pay attention to the fact that the parameters $v$, $\psi$, and ${\bf a}_\phi$, also depend on ${\bf p}$. Therefore, by eq.~(\ref{eq:dvdp}), we can show that
\begin{equation}
\frac{\partial {\bf m}_p^T}{\partial{\bf p}} =\left(v_0^2{\bf I} - \frac{1}{v_0^2}\frac{\partial  v_0}{\partial{\bf p}}\otimes\frac{\partial  v_0}{\partial{\bf p}} + \frac{1}{v_0}\frac{\partial^2 v_0}{\partial{\bf p}\partial{\bf p}^T}\right)\, ,
\label{formatensorialvelocity}
\end{equation}
where, as already mentioned, the symbol $\otimes$ represents the outer product operation. As seen in Appendix~\ref{Ap:devP}, we can write the centered-slowness second-order derivatives as
\begin{equation}
\begin{split}
\frac{\partial}{\partial p_j}\left(\frac{1}{v_0}\frac{\partial v_0}{\partial p_i}\right) &= \frac{1}{\cos^2\psi_0}\left(\bar{\bf f}_j\cdot\left(\frac{\partial^2{\cal H}_t}{\partial\bar{\bf w}\partial\bar{\bf w}^T} - v_0^2\bar{\bf I}\right)\bar{\bf f}_i\right. \\
&+ \left.2v_0^2\left(\bar{\bf f}_j\cdot\hat{\bf n}\right)\left(\bar{\bf f}_i\cdot\hat{\bf n}\right)\right)\, .  
\end{split}
\label{apEeq}
\end{equation}
Furthermore, in the case of
\begin{equation}
\frac{\partial {\bf m}_p^T}{\partial{\bf p}} = v_0^2{\bf I}\, ,    
\end{equation}
we say that the wavefront is isotropic in that propagation direction. In addition, the wave-propagation metric tensor \citep{Klime1994} in isotropic case,  where $\psi=0$, is
\begin{equation}
\frac{\partial^2{\cal H}_t}{\partial\bar{\bf w}\partial\bar{\bf w}^T} = v_0^2\bar{\bf I}\, .    
\end{equation}
Besides, we named eq.~(\ref{formatensorialvelocity}) as the reduced wave-propagation metric tensor due to the wave-propagation metric tensor nomenclature given by \cite{Klime:2002}. 

Also, using that one explicit equation, we can derive the direct expression for the other sub-matrices defined in the central-ray direction. Through forward mathematical manipulations, it is possible to show that 
\begin{equation}
\frac{\partial {\bf m}_q^T}{\partial{\bf q}} = \frac{1}{v_0}\frac{\partial^2 v_0}{\partial{\bf q}\partial{\bf q}^T}\, , \label{heterogeneidade}
\end{equation} 
where, by chain rule in eq.~(\ref{eq:dvdx}), we have that
\begin{equation}
\frac{\partial^2 v_0}{\partial q_i \partial q_j} = \hat{\bf e}_i\cdot\frac{\partial^2 v}{\partial\bar{\bf x}\partial\bar{\bf x}^T}\hat{\bf e}_j\, , \quad i,j=1,2.    
\end{equation}

Now, we compute the mixed expressions using the same procedure before, we get the following identities
\begin{equation}
\frac{\partial {\bf m}_p^T}{\partial{\bf q}} = \left(\frac{\partial {\bf m}_q^T}{\partial{\bf p}}\right)^T =\frac{1}{v_0^2}\frac{\partial v_0}{\partial {\bf q}}\otimes\frac{\partial v_0}{\partial {\bf p}}\, .
\label{mixed}
\end{equation}
We can observe the symplectic form of the propagation matrix in a sense established in \cite{Cerveny:2001}. Therefore, by Liouville's theorem, the system is conservative along the central ray trajectories. As shown in \cite{Cerveny:1972}, the formulation of the dynamic ray tracing system in terms of the Cartesian coordinate system ensures the symmetry between the matrices obtained with the mixed derivatives in ${\bf p}$ and {\bf q}. Consequently, such property is inherited when we formulate the problem in the ray-centered coordinate system, as shown in \cite{Klime1994} and also in \cite{Cerveny:2007}. Finally, with the expressions given explicitly, our formulation allows a qualitative understanding of the propagation phenomenon without the need for coordinate transformations for the $6\times 6$ Cartesian coordinate system.

\subsection{On initial Conditions}

The initial conditions for dynamic ray tracing are necessary for using the explicitly formulated Hamiltonian system. For such a task, consider the position and slowness vectors, starting from the origin, i.e., $t=0$, which applied in $\bar{\boldsymbol{\gamma}}_0 = [\gamma_1,\gamma_2, 0]^T$, in a general form as follows
\begin{equation}
{\bf q}_0(\boldsymbol{\gamma}) = {\bf q}(\bar{\boldsymbol{\gamma}}_0) \qquad \text{and} \qquad
{\bf p}_0(\boldsymbol{\gamma}) = {\bf p}(\bar{\boldsymbol{\gamma}}_0)\, .
\end{equation}
On the initial surface, the free parameter $\boldsymbol{\gamma}=[\gamma_1,\gamma_2]^T$ changes in accordance with the adopted initial conditions. Therefore, the initial conditions lead to
\begin{equation}
{\bf Q}_0 = \frac{\partial {\bf q}_0^T}{\partial \boldsymbol{\gamma}} \qquad \text{and} \qquad
{\bf P}_0 = \frac{\partial {\bf p}_0^T}{\partial \boldsymbol{\gamma}}\, .
\label{initialdef}
\end{equation}
It is important to note from eq.~(\ref{eq:dqdt_dqdgamma}) the parameter $\boldsymbol{\gamma}$ directly influences the ray-centered coordinates. Here, two appropriate initial conditions are presented in detail, namely that of a point source and that of the explosive reflector.

\begin{figure*}
\centering
\includegraphics[width=0.40\textwidth]{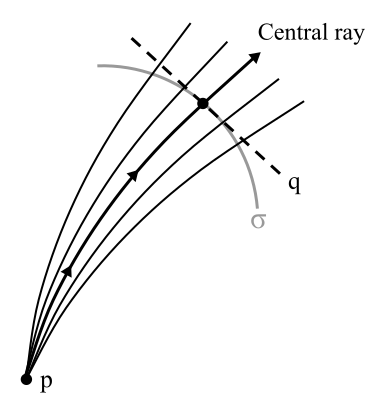}
\caption{A 2D schematic with the central and paraxial rays are emerging from a point source varying $p$ with a wavefront $\sigma$, in gray, and its tangent plane wave, with coordinate $q$, represented by the dashed line. }
\label{fig:feixe}
\end{figure*}

We start with the case of a point source. We take the parameter $\boldsymbol{\gamma} = {\bf p}$, and as all rays start from the same initial point and using the equations (\ref{initialdef}), we immediately get
\begin{equation}
{\bf Q}_0 ={\bf O}\qquad \text{and} \qquad
{\bf P}_0= {\bf I}.
\label{eq:pontualsource}
\end{equation}
These initial conditions are called the normalized point-source initial conditions, illustrated in Figure~\ref{fig:feixe}.

To consider a reflective surface source represented by the $\sigma$ function, it is possible to introduce an initial condition based on the curvature surface that osculates the wavefront propagation. For such construction, we assume that the initial surface is the wavefront surface from a point source started at some specific time $t<0$ but measured in $t=0$. For that, we choose precisely the centered-ray coordinate so that $\boldsymbol{\gamma} = {\bf q}$, which implies immediately from the equation (\ref{initialdef}) the following relations
\begin{equation}
{\bf Q}_0 = {\bf I}\qquad \text{and}\qquad
  {\bf P}_0 = \frac{\partial {\bf p}_0^T}{\partial {\bf q}}. 
\label{initialdefreflect}
\end{equation}

To establish explicit expressions for the initial conditions in eq.~(\ref{initialdefreflect}), we consider the following position parametrization $t({\bf q}) = \tau(\bar{\bf x}(\sigma({\bf q})))$, which implies in
\begin{equation}
{\bf p} = \frac{\partial t}{\partial{\bf q}} = \frac{\partial\tau}{\partial \sigma}\frac{\partial \sigma}{\partial{\bf q}}\, ,
\end{equation}
and  in eq.~(\ref{initialdefreflect}), yields 
\begin{equation}
{\bf P}_0 =\frac{\partial {\bf p}_0^T}{\partial {\bf q}} = \frac{\partial\tau}{\partial \sigma}\frac{\partial^2 \sigma}{\partial{\bf q}\partial{\bf q}^T} + \frac{\partial^2\tau}{\partial \sigma^2}\frac{\partial \sigma}{\partial{\bf q}}\otimes\frac{\partial \sigma}{\partial{\bf q}}\, .
\end{equation}
Therefore, the traveltime variation with respect to the vicinity of the central ray is adequately characterized by 
\begin{equation}
\left.\frac{\partial\tau}{\partial \sigma}\right|_{{\bf q}={\bf 0}} = \frac{\partial\tau}{\partial\bar{\bf r}}\cdot\frac{d\bar{\bf r}}{d\sigma} = \bar{\bf w}\cdot{\bf \hat{n}} = \frac{1}{v_0} .
\end{equation}
Therefore, the wavefront surface function $\sigma(t=0,{\bf q})$  represents an approximation of the wavefront in the vicinity of the central ray and, as a consequence, it implies that 
\begin{equation}
\left.\frac{\partial \sigma}{\partial {\bf q}}\right|_{{\bf q}={\bf 0}} = {\bf 0}\, , \quad \text{and} \quad
\left.\frac{\partial^2 \sigma}{\partial{\bf q}\partial{\bf q}^T}\right|_{{\bf q}={\bf 0}} = {\bf K}_0\, ,
\label{eq:d2sigmadq2}
\end{equation}
where ${\bf K}_0$ represents the surface curvature of the start point as shown in Figure~\ref{fig:5}. Note that ${\bf K}_0$ matches the reflector's curvature locally, i.e., the curvature matrix is measured on the plane generated at $\sigma_0=\sigma(t=0)$ with the coordinates given by the vector ${\bf q}$ at ${\bf 0}$. Finally, the explicit expressions for the initial condition are given as follows
\begin{equation}
{\bf Q}_0 = {\bf I}\qquad \text{and}\qquad
{\bf P}_0 =\frac{1}{v_0}{\bf K}_0. 
\label{initialdefreflectdeduzido}
\end{equation}
If the reflector is a plane, we have that ${\bf K}_0$ is the null matrix, and, therefore, the initial conditions becomes
\begin{equation}
{\bf Q}_0 = {\bf I}\qquad \text{and}\qquad
{\bf P}_0 = {\bf O}\, . 
\label{eq:init_plane}
\end{equation} 
These initial conditions are called the normalized telescopic point or normalized plane wavefront initial conditions.

\section{Physical interpretation of the emerging wavefronts and application}

In this section, we show two applications of the explicit formulation of the Hamiltonian in physical attributes to traveltime parameters. First, we provide interpretations of the physical attributes of the wavefront measured at the surface. Second, we present the exact velocity of the NMO derived directly from the proposed formulation. To clarify the examples, we suppose the medium is homogeneous for all models.

Dynamic ray tracing describes the wavefront curvature evolution along a central ray. Based on eq.~(\ref{eq:d2sigmadq2}), we can formulate this evolution, running at $t_0$, as
\begin{equation}
\left.\frac{\partial^2 \sigma}{\partial{\bf q}\partial{\bf q}^T}\right|_{{\bf q}={\bf 0}} = v_0{\bf P}{\bf Q}^{-1} = {\bf K}_{t_0}\, ,     
\end{equation}
where ${\bf K}_{t_0}$ is a wavefront curvature matrix. Therefore, as already noted earlier, the initial conditions influence the shape of the wavefront propagation in the medium, and consequently, it controls the curvature of the wavefront. Besides, as demonstrated by \cite{Hubral:1980a}, we can obtain crucial kinematic information for the inversion process by understanding the wavefront behavior along the raypath propagation. We can create two imaginary wavefronts by changing the initial conditions for ${\bf Q}$ and ${\bf P}$. First, the normal-incidence-point (NIP) wavefront is caused by a  point source at the NIP on the reflector. Second, the Normal wavefront is generated by propagating the reflector's local curvature. Besides, these imaginary wavefronts share the same normal-incidence ray with the corresponding slowness vector. Here, we analyze these wavefronts for homogeneous anisotropic media in a straightforward way. It is important to point out here that the normal incidence refers to the phase vector on the surface.

\begin{figure*}
\centering
\includegraphics[width=0.65\textwidth]{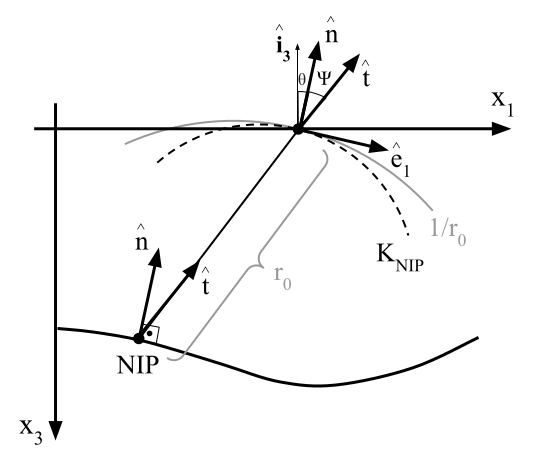}
\caption{A 2D schematic representation of a NIP wavefront departing from a reflector in an anisotropic medium. Note that, due to anisotropy, the departing angle of the ray from the NIP is not normal with the reflector's surface.} 
\label{fig:4}
\end{figure*}
\begin{figure*}
\centering
\includegraphics[width=0.65\textwidth]{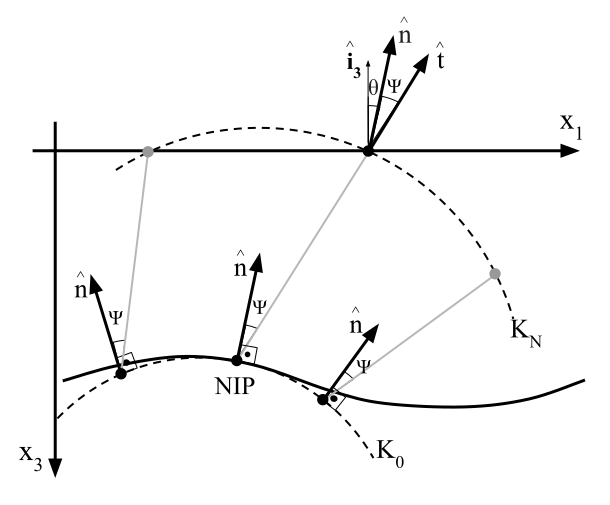}
\caption{A 2D schematic representation of a Normal wavefront departing from an osculating surface at the reflector, with curvature $K_0$, in an anisotropic medium. Note that, due to anisotropy, the departing angle of the ray from the Normal wavefront is not normal with the reflector's surface.}
\label{fig:5}
\end{figure*}

In order to analyze these imaginary wavefronts, we examined their curvature matrices, ${\bf K}_\text{NIP}$ and ${\bf K}_\text{N}$, for the NIP and Normal wavefronts, respectively. Figures~\ref{fig:4} and~\ref{fig:5} illustrate both wavefronts propagating on an anisotropic homogeneous medium. Taking the normalized point-source initial conditions (see, eq.~(\ref{eq:pontualsource})) and solving the Hamiltonian system given by eq.~(\ref{seconsystem}), we have the expression lead to
\begin{equation}
{\bf K}_\text{NIP} = \frac{1}{r_0}\left[\frac{{\bf C}_0}{v_0^2}\right]^{-1}\, ,  
\label{eq:KNIP}
\end{equation}
where $r_0 = t_0 v_0$ is the radius of curvature of a spherical wavefront centered in the NIP, and 
\begin{equation}
 {\bf C}_0 = \frac{\partial {\bf m}_p^T}{\partial{\bf p}}\, ,   
\end{equation}
as the reduced wave-propagation metric tensor in the central-ray direction. The anisotropic-stretching factor ${\bf C}_0/v_0^2$ influences the curvature of the NIP source wavefront, making it non-spherical. Furthermore, from eq.~(\ref{eq:geospreading}), we have the geometrical spreading on the NIP central ray as
\begin{equation}
 \Lambda_\text{NIP} = v_0\sqrt{\cos\psi_0|\det{\bf K}_\text{NIP}|^{-1}}\, .   
\end{equation}
It is important to mention that $\Lambda_\text{NIP}$ is the geometric spreading by a point source.

To analyze the behavior of the curvature of a wavefront emitted by an explosive reflector, we consider a source that matches the reflector curvature at the NIP position, as shown in Figure~\ref{fig:5}. The initial conditions are given by eq.~(\ref{initialdefreflectdeduzido}) starting in the NIP source. Using the Hamiltonian system to calculate ${\bf Q}$ and ${\bf P}$ implies that
\begin{equation}
{\bf Q}(t_0)=\frac{t_0}{v_0}{\bf C}_{0}{\bf K}_0 + {\bf I}\, , \quad {\bf P}(t_0) = \frac{1}{v_0}{\bf K}_0\, .
\end{equation} 
Combining the equations one together eq.~(\ref{eq:KNIP}), yields
\begin{equation}
{\bf K}_\text{N} = {\bf K}_0\left[{\bf K}_\text{NIP}^{-1}{\bf K}_0 + {\bf I}\right]^{-1}\, .
\label{curvaturageral}
\end{equation}
From that, we can describe the geometrical spreading in the central-ray direction from an explosive reflector as
\begin{equation}
 \Lambda_\text{N} = \sqrt{\cos\psi_0|\det\left[{\bf K}_\text{NIP}^{-1}{\bf K}_0 + {\bf I}\right]|}\, .   
\end{equation}
Note that if ${\bf K}_\text{NIP} = -{\bf K}_0$, then we have a caustic situation implied in $\Lambda_\text{N} = 0$. Also, in the case of a plane reflector, we have $ \Lambda_\text{N} = \sqrt{\cos\psi_0}$, where we can observe that only the anisotropic deviation influences the geometric spreading in such a situation. 

In summary, the NIP theorem shows that the reflector's shape has no influence on the ${\bf K}_\text{NIP}$ \citep{chernjak:1979, Hubral:1980}. Therefore, in the homogeneous media, the NIP wavefront curvature only carries information about the anisotropy of the medium, and the anisotropic-stretching factor determines such an influence. For the case of normal curvature, in the homogeneous media, we observe that the reflector's curvature is modified by its geometrical spreading, i.e., ${\bf K}_\text{N} = {\bf K}_0{\bf Q}^{-1}$. Therefore, any analysis of ${\bf K}_\text{N}$ enables us to understand the shape of the reflector as long as we have the information about ${\bf K}_\text{NIP}$.

\subsection{Exact expression for normal moveout velocity}

In order to understand the hyperbolic moveout velocity \citep{ALCHALABI:1973}, we analyze its theoretical version, known as NMO velocity. We show that the proposed formulation can derive the NMO velocity from a common-midpoint (CMP) configuration related to the dip plane of a reflector overburdened by a homogeneous anisotropic layer. We present the matrix version of the NMO velocity in terms of dynamical ray-paraxial formulation.   

By definition, the two-way CMP traveltime approximation ${\cal T}_\text{CMP}$ is given by
\begin{equation}
{\cal T}_\text{CMP}({\bf x}_h, {\bf x}_0)^2 = \tau_{0}^2 + 4\Delta{\bf x}_h\cdot\left[{\bf V}_\text{NMO}\right]^{-2}\Delta{\bf x}_h\, ,
\label{traveltime_1}
\end{equation}
where $\tau_0$ is two-way traveltime from the NIP to ${\bf x}_0$ on the measurement surface, $\Delta{\bf x}_h = {\bf x}_h - {\bf 0}$, ${\bf x}_h = ({\bf x}_r - {\bf x}_s)/2$ is the  source-receiver half-offset vector position, ${\bf x}_0 = ({\bf x}_r + {\bf x}_r)/2$ is the reference midpoint position, and ${\bf x}_s$ and ${\bf x}_r$ are the source and receiver positions on the measurement surface, respectively. Therefore, using the solution of the Hamiltonian system, it is possible to interpret the kinematic parameters, ${\bf V}_\text{NMO}$, related to the ray-paraxial wavefront. 

Based on the work of \cite{Hubral:1980}, we can show the expression for ${\bf V}_\text{NMO}$, i.e., the NMO-velocity matrix, as
\begin{equation}
4\left[{\bf V}_\text{NMO}\right]^{-2} = \tau_0\left.\frac{\partial^2 \tau}{\partial {\bf x}_h\partial{\bf x}_h^T}\right|_{({\bf 0},{\bf x}_0)}\, , 
\end{equation}
where using the ray-paraxial approach, we can show that
\begin{equation}
\frac{\partial^2 \tau}{\partial {\bf x}_h\partial{\bf x}_h^T} = \frac{2}{v_0}\left(\frac{\partial{\bf q}^T}{\partial{\bf x}_h}\right)^T{\bf K}_\text{NIP}\left(\frac{\partial{\bf q}^T}{\partial{\bf x}_h}\right)\, . 
\label{eq:d2tdx2}
\end{equation}
Now, setting ${\bf x} = x_1\hat{\bf i}_1 + x_2\hat{\bf i}_2$, we can make the following derivative
\begin{equation}
\frac{\partial{\bf q}^T}{\partial{\bf x}_h} = \frac{\partial{\bf q}^T}{\partial{\bf x}}\frac{\partial{\bf x}^T}{\partial{\bf x}_h}\, ,
\end{equation}
where
\begin{equation}
\begin{split}
\frac{\partial{\bf x}^T}{\partial{\bf x}_h} =\ & 2{\bf I}\, , \\
\frac{\partial q_i}{\partial x_j} =\ & \frac{1}{\cos\psi}\left(\hat{\bf i}_j\cdot\bar{\bf f}_i\right)\, , \ i,j=1,2\, .
\end{split}
\end{equation}

The approach described above can be used to obtain the three-dimensional version of ${\bf V}_\text{NMO}$. In this case, considering the Cartesian coordinates, we have the following expression 
\begin{equation}
\left[{\bf V}_\text{NMO}\right]^2 = v_0^2{\bf H}\left[\frac{{\bf C}_0}{v_0^2}\right]{\bf H}^{T}\, ,   
\end{equation}
where ${\bf H}$ is an transformation matrix given by
\begin{equation}
{\bf H} = \left(\frac{\partial{\bf q}^T}{\partial{\bf x}}\right)^{-1}\, .    
\end{equation}
Thus, we obtain a link between the Cartesian and ray-centered formulation via the ${\bf H}$ matrix. Therefore, by making a coordinate conversion, it is possible to get the exact expression of the NMO velocity after solving the three-dimensional version of the dynamic ray-tracing system. Furthermore, in a homogeneous model, the NMO velocity estimated from reflection traveltimes recorded from a CMP geometry provides valuable information about the subsurface's velocity field and anisotropic parameters. Finally, we show that there is a relationship between the measured NMO velocity and the anisotropic-stretching factor matrix.

\section{Velocity spreading factor in general anisotropic media}

This section presents the relationship between the Dix velocity and the time-migration curvature matrices. The time-migration traveltime is based on a diffraction traveltime approximation equation that, commonly, is a function of the source and receiver coordinates. An underlying assumption of the procedure is that this traveltime approximation can osculate the diffraction traveltime at its apex point. The starting point is the following approximation of the traveltime equation as
\begin{equation}
\begin{split}
{\cal T}_{M}({\bf x}_s,{\bf x}_r) &= \sqrt{t_0^2 + \Delta{\bf x}_s\cdot[t_0{\bf \Gamma}_M]\Delta{\bf x}_s} \\
&+ \sqrt{t_0^2 + \Delta{\bf x}_r\cdot[t_0{\bf \Gamma}_M]\Delta{\bf x}_r}\, ,   
\end{split}
\end{equation}
where $\Delta {\bf x}_s = {\bf x}_0 - {\bf x}_s$, $\Delta {\bf x}_r = {\bf x}_0 - {\bf x}_r$ on measurement plane, and setting the traveltime $t$ as in eq.~(\ref{eq:t_q}) defines 
\begin{equation}
 {\bf \Gamma}_M = \left.\frac{\partial^2 t}{\partial{\bf q}\partial{\bf q}^T}\right|_{{\bf q}={\bf 0}}\, .  
\end{equation}

\begin{figure*}
\centering
\includegraphics[width=1\textwidth]{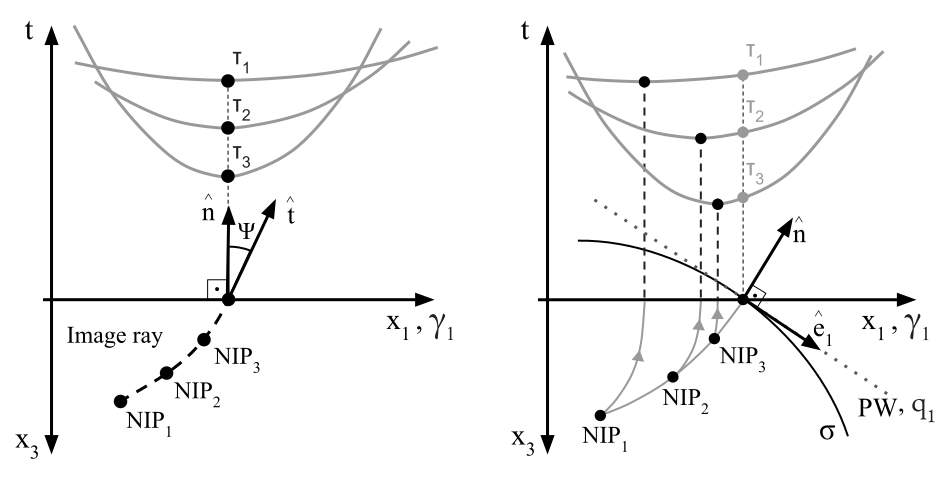}
\caption{Left: Traveltime responses, represented by $\tau_1$, $\tau_2$, and $\tau_3$, related to point sources, NIP$_1$, NIP$_2$, and NIP$_3$, respectively. In this case, all the diffraction traveltimes are aligned with the image ray, where the phase vector arrives perpendicular to $x_1$. Right: Similar to the previous situation, only now, the image ray is in relation to the referential plane PW, which did not arrive with normal incidence with respect to $x_1$. From the NIP$_1$, NIP$_2$, and NIP$_3$ are also the image rays in relation to the measurement plane.}
\label{fig:7}
\end{figure*}

The relationship between wavefront propagation velocity and time-migration velocity has been well-established in isotropic media.  As shown in \cite{Cameron:2007}, there is a relationship between the time-migration slowness-squared matrix 
\begin{equation}
\left[{\bf V}_M\right]^{-2} = t_0{\bf \Gamma}_{M} = \frac{t_0}{v_0}{\bf K}_M\, ,    
\end{equation}
and the Dix velocity \citep{Dix:1955} given by
\begin{equation}
\left[{\bf V}_{\text{Dix}}\right]^{2}=\frac{\partial}{\partial t_0}{\bf \Gamma}_{M}^{-1}(t_0)\, ,
\label{dix}
\end{equation}
where ${\bf K}_{M}$ is the curvature matrix of a NIP wavefront on an image ray \citep{HUBRAL:1977}. Figure~\ref{fig:7} (left side) shows an image ray crossing the NIP positions, and each one of them is represented in time by the traveltimes $\tau$ over the same $\gamma_1$. At such points in the time-migrated domain, the time-migration velocities are extracted to perform the process described by eq.~(\ref{dix}).

The image-ray concept explains how a depth velocity model can be converted to time coordinates in an isotropic medium. For our purpose, we define an image ray as the raypath of the central ray whose phase vector arrives normal to the measurement surface. The main physical idea about the raypath of the image ray can be seen in the part of Lagrangian mechanics already presented in section~\ref{sec:3}. In general anisotropic media, we investigate this same relation through phase velocity, and therefore, the framework established in \cite{Cameron:2007} can perfectly adapt to our formulation.

Taking into account the time-dependence of the ${\bf K}_M$, we set ${\bf M} = v_0{\bf K}_M^{-1}$,  which implies ${\bf M}={\bf Q}{\bf P}^{-1}$. Therefore, it should be noted that dynamic equations are time-dependent and can be written as 
\begin{equation}
\frac{d}{d t} \left[\begin{matrix}
\mathbf{Q}\\
\mathbf{ P}
\end{matrix}\right]=\left[\begin{matrix}
{\bf Y}_0^T &{\bf C}_0 \\
-\mathbf{V}_{0} & -{\bf Y}_0
\end{matrix}\right]  \left[\begin{matrix}
\mathbf{Q}\\
\mathbf{P}
\end{matrix}\right]\, ,
\label{sisheteraniso}
\end{equation} 
where 
\begin{equation}
{\bf Y}_{0}= \frac{1}{v_0^2}\frac{\partial v_0}{\partial {\bf q}}\otimes\frac{\partial v_0}{\partial {\bf p}}\, , \quad {\bf V}_{0} = \frac{1}{v_0}\frac{\partial^2 v_0}{\partial{\bf q}\partial{\bf q}^T}\, ,
\end{equation}
with all matrix elements evaluated in the central-ray direction. Consequently, the system of eqs.~(\ref{sisheteraniso}) allows a complete understanding of the wave propagation phenomenon and the velocity spreading factor in a general medium. Now, written in terms of ${\bf Q}$ and ${\bf P}$ as a function of $\bar{\boldsymbol{\gamma}}$ on an image-ray, we have
\begin{equation}
    {\bf M}\left(\bar{\boldsymbol{\gamma}}\right)={\bf Q}(\bar{\boldsymbol{\gamma}}){\bf P}^{-1}(\bar{\boldsymbol{\gamma}})\, .
    \label{definv}
\end{equation}
 
Note that, based on eq.~(\ref{dix}), we can obtain the variation in $t_0$ of eq.~(\ref{definv}) as
\begin{equation}
\frac{\partial {\bf M} }{\partial t_0} = {\bf Q}_v{\bf P}^{-1} - {\bf Q}{\bf P}^{-1}{\bf P}_v{\bf P}^{-1}\, ,
\label{taylorcameron}
\end{equation}
where, for convenience, we introduce the notation
  \begin{equation}
{\bf Q}_v = \frac{\partial{\bf Q}}{\partial t_0}\, , \quad  {\bf P}_v = \frac{\partial{\bf P}}{\partial t_0}\, . 
\label{qvaria}
\end{equation} 

Following \cite{Cerveny:2001}, we define the matrices ${\bf Q}_2$ and ${\bf P}_2$ as the solutions of the dynamic ray tracing system described by eq.~(\ref{sisheteraniso}) when a depth point source is considered, i.e., the system has initial conditions described according to eq.~(\ref{eq:pontualsource}). Likewise, we define by ${\bf Q}_1$ and ${\bf P}_1$, which assume a family of rays in the vicinity of a central ray known as a normalized telescopic point in the subsurface, i.e., the system has initial conditions described according to eq.~(\ref{eq:init_plane}).

Now, for the characterization of the initial conditions for this problem, we indicate ${\bf X}_q = {\bf Q}_2$ and ${\bf X}_p = {\bf P}_2$ in eq.~(\ref{taylorcameron}).  Therefore, using such a relation, we make the following derivation
\begin{equation}
\frac{d{\bf X}_j}{dt} = \frac{\partial {\bf X}_j}{\partial t_0}\frac{dt_0}{dt} + \sum_{i=1}^2\frac{\partial {\bf X}_j }{\partial \gamma_i}\frac{d\gamma_i}{dt}\, , \quad j=p,q\, ,
\end{equation}
which in $t=0$, yields 
\begin{equation}
\frac{dt_0}{dt} = 1\, , \quad \frac{d\boldsymbol{\gamma}}{dt} = {\bf 0}\, .    
\end{equation}
From that, we have the following result
\begin{equation}
\frac{d{\bf X}_j}{dt} =  \frac{\partial {\bf X}_j}{\partial t_0}\, .    
\end{equation}
Therefore, one can find the initial conditions, namely
\begin{equation}
{\bf Q}_{v0} = \left.\frac{\partial{\bf Q}_2}{\partial t_0}\right|_0 = {\bf C}_0\, , \quad \text{and} \quad {\bf P}_{v0} = \left.\frac{\partial{\bf P}_2}{\partial t_0}\right|_0 = -{\bf Y}_{0}\, .    
\label{qvaria_ini}
\end{equation}
\begin{figure*}
\centering
\includegraphics[width=0.5\textwidth]{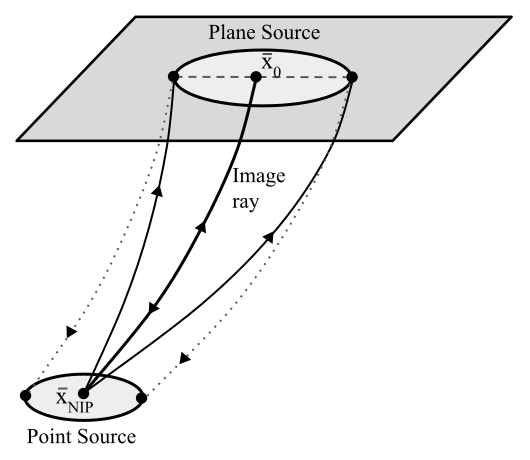}
\caption{Two scenarios with two different kinds of sources in a ray tube. First, representing a point source, at NIP, from a ray flow propagating in the direction of the measurement surface. Secondly, represents a line source propagating into the depth direction, with the paraxial rays representing the telescopic family.}
\label{fig:6}
\end{figure*}

From the expression (\ref{sisheteraniso}), it is possible to obtain a system in terms of eqs.~(\ref{qvaria}) as follows
\begin{equation}
\frac{d}{d t}\left[\begin{matrix}
{\bf Q}_v\\
{\bf P}_v
\end{matrix}\right]=\left[\begin{matrix}
{\bf Y}_0^T & {\bf C}_0 \\
-{\bf V}_{0} & -{\bf Y}_0
\end{matrix}\right] \left[\begin{matrix}
{\bf Q}_v\\
{\bf P}_v
\end{matrix}\right]\, ,
\end{equation} 
with initial conditions given by eqs.~(\ref{qvaria_ini}). Note that its solutions combined with eq.~(\ref{taylorcameron}) allow us to obtain an expression for the Dix velocity considering the propagated wave in relation to a point source in depth. Also, there is no difficulty in showing that
\begin{equation}
\left[\begin{matrix}
{\bf Q}_v\\
{\bf P}_v
\end{matrix}\right]=\left[\begin{matrix}
{\bf Q}_1 & {\bf Q}_2 \\
{\bf P}_1 & {\bf P}_2
\end{matrix}\right]\left[\begin{matrix}
{\bf C}_0\\
-{\bf Y}_0
\end{matrix}\right]\, ,
\end{equation} 
is the solution for the system with the given initial conditions. Finally, replacing the answer in eq.~(\ref{taylorcameron}) and using the condition well known that the propagation matrix is symplectic \citep{Cerveny:2001,Popov:2002}, it follows that
\begin{equation}
\frac{\partial {\bf M} }{\partial t_0} = \left({\bf Q}_1 - {\bf Q}_2{\bf P}_2^{-1}{\bf P}_1\right){\bf C}_0{\bf P}_2^{-1} = {\bf P}_2^{-T}{\bf C}_0{\bf P}_2^{-1}\,.   
\end{equation}

Considering the propagation matrix theory described in \cite{Cerveny:1972} and following the detailed demonstration in \cite{Cameron:2007}, that can be adapted  for our formulation, there exists a reciprocity argument that ensures the relation ${\bf P}^{T}_{2}(\bar{\bf x}_\text{NIP},\bar{\bf x}_0)={\bf Q}_{1}(\bar{\bf x}_0,\bar{\bf x}_\text{NIP})$ for two points $\bar{\bf x}_\text{NIP}$ and $\bar{\bf x}_0$. In our case, $\bar{\bf x}_\text{NIP}$ is the NIP in depth and $\bar{\bf x}_0$ a point at measurement surface and such reciprocity property implies that a propagation under initial conditions given by the "normalized telescope point" located at the surface has the same form, at depth point, that the solution of a point source located at depth and evaluated at the surface. Such feature is illustrated in Figure~\ref{fig:6} and, as a consequence, it follows that
\begin{equation}
\frac{\partial {\bf M}}{\partial t_0}\ = {\bf Q}_{1}^{-1}{\bf C}_0{\bf Q}_{1}^{-T} =  v_0^2\left[{\bf Q}_{1}^{-1}\frac{{\bf C}_0}{v_0^2}{\bf Q}_{1}^{-T}\right]\,.
\label{chavespread}
\end{equation}
Based on these results, different from \cite{Iversen:2008}, we define the velocity-spreading factor ${\bf Q}_F$ as
\begin{equation}
[{\bf Q}_F]^{2} = {\bf Q}_{1}^{T}\left[\frac{{\bf C}_0}{v_0^2}\right]^{-1}{\bf Q}_{1}\, .    
\end{equation}
The above expression can represent the time-velocity matrix ${\bf V}_{\text{Dix}}$ computed from eq.~(\ref{dix}) leading to
\begin{equation}
{\bf V}^{2}_{\text{Dix}} ={\bf Q}_{1}^{-1}{\bf C}_0{\bf Q}_{1}^{-T} = v_0^2\left[{\bf Q}_F\right]^{-2}\, .
\label{anisodix}
\end{equation}
It is necessary to take care that the raypath starts from the measurement surface and goes towards the NIP, using the normalized telescopic point initial conditions.

An alternative approach to recover the direction-depended velocity field of the underlying effective medium is to start with a generalized time migration velocity matrix. Consider that in a homogeneous neighborhood around the measurement plane that is thin enough to the spatial derivatives of the phase velocity are null over the measurement points $\bar{{\bf x}}_0$. Therefore, in analogy with the image ray case, a generalized time migration velocity matrix can be defined as
\begin{equation}
[{\bf V}_{\text{M}}(t_0)]^2 = {\bf H}^{-1}[{\bf V}_{\text{NMO}}(t_0)]^2{\bf H}^{-T}\, .    
\end{equation}
The ${\bf H}$ matrix depends only on the phase vector orientation parameter while ${\bf V}_\text{Dix}$ and ${\bf V}_\text{NMO}$ are time dependents. As seen in the right part of Figure~\ref{fig:7}, we can observe over the time axis in $\bar{\boldsymbol{\gamma}}$ coordinate, we pick the NMO velocities which have the same slowness direction, i.e., the same traveltime slopes parameters. Therefore, after the coordinate changing by the inverse of the ${\bf H}$ matrix, it is possible to apply the Dix procedure, as in eq.~(\ref{dix}), to obtain the Dix velocity matrix for that direction. This concept of creating Dix velocities via NMO velocities through a transform matrix has already been explored by \cite{Gelius:2015}. Finally, to obtain the Dix velocity matrix, which also depends on the direction of arrival of the phase vector, we can use \cite{Coimbra:2019} by diffraction separation as a practical way to extract the slope vector and velocity matrix for any phase direction. 

\section{Eikonal-type equation for time migration}

\begin{figure*}
\centering
\includegraphics[width=1.0\textwidth]{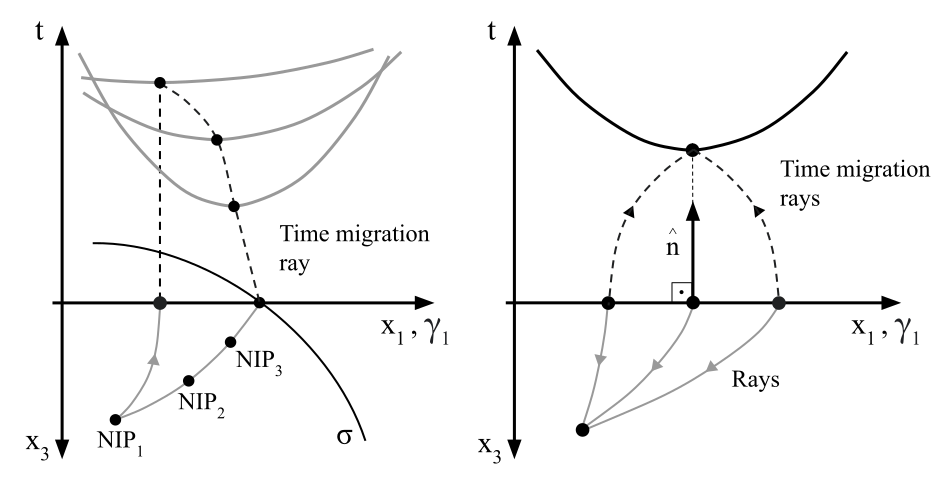}
\caption{The trajectories of time-migration rays in the time-migrated domain.}
\label{fig:9}
\end{figure*}

In this section, we derive the Eikonal-type equation for time migration, namely the one that determines the time-migration rays \citep{Fomel:2021}. We describe the time migration rays as the raypaths in the time domain where each point on the raypath in depth is taken to the apex of its traveltime response, assuming that such a point is a scattering point, as seen in the left side from Figure~\ref{fig:9}. 

We start with the following relation
\begin{equation}
t_0(\bar{\boldsymbol{\gamma}}) = t\, ,
\label{ansatz_mod2}
\end{equation}
where $\bar{\boldsymbol{\gamma}} = [\gamma_1,\gamma_2,t]^T$ and recast in the form
\begin{equation}
t_0(\bar{\boldsymbol{\gamma}}) = t(\bar{\bf q}(\bar{\boldsymbol{\gamma}}))\, ,
\label{ansatz_mod2_ap}
\end{equation}
where $t_0(\bar{\boldsymbol{\gamma}})$ is the solution of the Eikonal-type equation for a general medium with running parameter $t$. In other terms, eq.~(\ref{ansatz_mod2}) represents the wavefront of a time migration ray at the instant $t$. Therefore, applying to both sides of eq.~(\ref{ansatz_mod2_ap}) the partial derivatives concerning $\bar{\boldsymbol{\gamma}}$, we find
\begin{equation}
\frac{\partial t_0}{\partial\bar{\boldsymbol{\gamma}}} = \frac{\partial\bar{\bf q}^T}{\partial\bar{\boldsymbol{\gamma}}}\frac{\partial t}{\partial\bar{\bf q}}\, . 
\label{part_der1}
\end{equation}
Introducing the Jacobian matrix
\begin{equation}
\bar{\bf Q} = \frac{\partial\bar{\bf q}^T}{\partial\bar{\boldsymbol{\gamma}}}\, ,
\label{grads1}
\end{equation}
makes eq.~(\ref{part_der1}) can be recast in vector form as
\begin{equation}
\bar{\bf Q}^{-1}\bar{\bf g} = \bar{\bf p}\, ,
\label{grads2}
\end{equation}
where the time migration slowness-type vector is described as 
\begin{equation}
\bar{\bf g} = \frac{\partial t_0}{\partial\bar{\boldsymbol{\gamma}}}\, .    
\end{equation}
Left multiplication of eq.~(\ref{grads2}) by its conjugate equation, we readily obtain
\begin{equation}
\bar{\bf g}\cdot\bar{\boldsymbol{\Sigma}}\bar{\bf g} = \bar{\bf p}\cdot\bar{\bf p}=\frac{1}{v_{g}^{2}} \, ,
\label{grads4}
\end{equation}
where $\bar{\bf p}$ is given by (\ref{eq:pEg}) and, as consequence from eq.~(\ref{eq:dqdt_dqdgamma}), follows that
\begin{equation}
\bar{\boldsymbol{\Sigma}}  =  \left(\bar{\bf Q}^T \bar{\bf Q}\right)^{-1} 
 =  \begin{bmatrix}
\left({\bf Q}_1^T {\bf Q}_1\right)^{-1} & {\bf 0} \\
{\bf 0} & v_g^{-2}
\end{bmatrix} \, .
\label{grads5}
\end{equation}

Eq.~(\ref{grads4}) constitutes the Eikonal-type equation for the general case and, for the two-dimensional isotropic case, coincides with the Eikonal type from \cite{Fomel:2021}. However, given the validity conditions of the paraxial-ray theory, the deduction makes clear that such an equation is valid for any general media with anisotropy included.  
Taking the velocity dependent expression  $[{\bf V}_G(\bar{\boldsymbol{\gamma}})]^2 = v_g^2\left({\bf Q}_1^T {\bf Q}_1\right)^{-1}$ in eq.~(\ref{grads5}), we have the Hamiltonian as 
\begin{equation}
{\cal H}_{t_0}(\bar{\boldsymbol{\gamma}},\bar{\bf g}) = \left({\bf g}\cdot[{\bf V}_G(\bar{\boldsymbol{\gamma}})]^2{\bf g} + g_3^2 - 1\right)/2\,.
\label{eikonal_migration}
\end{equation}

In agreement with the Hamiltonian theory, the Eikonal-type equation that governs time-migration rays is determined by the characteristic curves \citep{Courant:1989} as follows
\begin{equation}
\begin{split}
\frac{d{\boldsymbol{\gamma}}}{dt} &= \lambda_t\frac{\partial{\cal H}_{t_0}}{\partial{\bf g}} \, ,\\
\frac{d\bar{\bf g}}{dt} &= -\lambda_t\frac{\partial{\cal H}_{t_0}}{\partial\bar{\boldsymbol{\gamma}}} \, , \\
\frac{dt_0}{dt} &= \left(\frac{\partial{\cal H}_{t_0}}{\partial g_3}\right)^{-1} = \lambda_t\, .
\end{split}
\end{equation}
Figure~\ref{fig:9} shows these time-migration rays in the time-migrated domain.

For our purpose, in the time-migrated domain, the wavefront propagation time $t$ is the running parameter of the time-migration ray. The ordinary differential system can be solved numerically when we impose appropriate initial conditions. Therefore, we start from the measurement plane towards the apex point to take advantage of certain simplifications (right side of Figure~\ref{fig:9}), that is $t(0)=0$, $\boldsymbol{\gamma}_0 = {\bf x}_0$, and 
\begin{equation}
{\bf g}_0 = \left.\frac{\partial \tau}{\partial {\bf x}}\right|_{{\bf x}={\bf x}_0}\, , \quad g_{3}(0) = \sqrt{1 - (\|{\bf g}_0\|v_g)^2} \, .   
\end{equation}
Observe that ${\bf g}_0$ is the slope vector of the diffraction traveltime at ${\bf x}_0$ on the measurement plane, and such a plane ${\bf Q}_1={\bf I}$, which implies in $[{\bf V}_G]^2=v_g^2{\bf I}$.

Therefore, in order for Dix velocity to be used to model time-migration rays, it is necessary that
\begin{equation}
{\bf V}_\text{Dix}={\bf V}_{G}\, . 
\label{compare}
\end{equation}
In other words, the relationship between the Dix velocity and the time-migration ray velocity is that the anisotropic-stretching factor is equal to the inverse anisotropic deviation squared, in mathematical form as
\begin{equation}
\frac{{\bf C}_0}{v_0^2} = \frac{1}{\cos\psi_0^2}{\bf I}\, .  
\label{eq:final}
\end{equation}
However, from eqs.~(\ref{formatensorialvelocity}), (\ref{tr1}), and~(\ref{tr2}), the relation in eq.~(\ref{eq:final}) is true if, and only if, 
\begin{equation}
 \frac{\partial^2{\cal H}_t}{\partial\bar{\bf w}\partial\bar{\bf w}^T} = v_0^2\bar{\bf I}\, ,   
\end{equation}
i.e., the medium is indeed isotropic in that phase direction. Therefore, as a theorem, we say that the necessary and sufficient condition for the Dix and time-migration ray velocities to be the same for all propagation directions is that the medium is isotropic.

\section{Conclusion}

In conclusion, a deep understanding of time-migration interval velocity and velocity-spreading factor is crucial for accurate time-to-depth conversion in seismic imaging. This work proposes a comprehensive framework that describes wavefront propagation explicitly in phase velocity, providing several advantages over existing literature. By doing so, we understand the contributions of anisotropy and heterogeneity to wave propagation and their effects on the wavefront. The proposed framework can be converted from Cartesian coordinates to ray-centered coordinates and applies to various seismic processing procedures. In addition, it generalizes the relationship between Dix velocities, defined in the time migration coordinate domain, and the physical velocity described in the depth domain, making it clear how anisotropy influences the geometric spreading factor. Significantly, even in a homogeneous anisotropic medium, the Dix velocity differs from the migration velocity, highlighting the need to consider anisotropy in seismic imaging. With the help of this framework, we can develop computational techniques to identify the effects of anisotropy on seismic data and distinguish it from the impact of medium heterogeneity. In future work, this solid understanding of the physical phenomenon will help us improve seismic imaging and increase our ability to interpret subsurface features accurately.

\begin{acknowledgments}
The authors thank the High-Performance Geophysics (HPG)
team for technical support. This work was possible thanks to
the support of Petrobras.
\end{acknowledgments}

\bibliographystyle{gji}
\bibliography{paper}

\appendix
\section{On Ray-velocity vector}
\label{Appendix:A}

In order to obtain a computable version of eq.~(\ref{system1}), we take 
\begin{equation}
\frac{\partial v}{\partial\bar{\bf w}} = \frac{\partial\hat{\bf n}^T}{\partial\bar{\bf w}}\frac{\partial v}{\partial\hat{\bf n}}\, ,  
\label{eq:dvdw}
\end{equation}
and, together from eq.~(\ref{eq:raypath}), we have
\begin{equation}
\hat{\bf n} = \frac{\bar{\bf w}}{\|\bar{\bf w}\|}\, ,    
\end{equation}
which applying the derivative with respect to $\bar{\bf w}$, we arrive at
\begin{equation}
\frac{\partial\hat{\bf n}^T}{\partial\bar{\bf w}} = \frac{1}{\|\bar{\bf w}\|}\left(\bar{\bf I} - \hat{\bf n}\otimes\hat{\bf n}\right)\, ,  
\label{eq:dndw}
\end{equation}
where the symbol $\otimes$ represents the outer product operation resulting, here, in a $3\times3 $ matrix. Substituting eq.~(\ref{eq:dndw}) into eq.~(\ref{eq:dvdw}), we get the derivative of the phase velocity with respect to the slowness vector as
\begin{equation}
\frac{1}{v}\frac{\partial v}{\partial\bar{\bf w}} = \frac{\partial v}{\partial\hat{\bf n}} - \left(\frac{\partial v}{\partial\hat{\bf n}}\cdot\hat{\bf n}\right)\hat{\bf n}\, .  
\label{eq:dvdw:dvdn}
\end{equation}

\section{Hamiltonian formula deduction}
\label{Appendix:Hamiltonian}

To show the Hamiltonian equation resulting from the Legendre transformation given by eq.~(\ref{hamilton}) in simplified coordinates of centered rays. We use the conjugate-type momenta in eq.~(\ref{lagrange}) as follows,
\begin{equation}
{\bf p}=\frac{\partial {\cal L}_s}{\partial \dot{\bf q}}\, .
\label{lagrange-slown}
\end{equation}
By using the explicit Lagrangian expression and taking into account the phase-velocity function $v$, we have that
\begin{equation}
 {\bf p}=\frac{1}{v}\frac{\sin\psi{\bf a}_\phi  + \dot{\bf q}}{\sqrt{\chi^2  + 2\sin\psi{\bf a}_\phi\cdot\dot{\bf q} + \dot{\bf q}\cdot\dot{\bf q}}}\, ,
 \label{lagrange-slow}
\end{equation}
and, now, we are going to solve the equation for $\dot{\bf q}$. Note that,
\begin{equation}
(vp_{i})^2=\frac{(a_i\sin\psi + \dot{q}_i)^2}{\chi^2  + 2\sin\psi{\bf a}_\phi\cdot\dot{\bf q} + \dot{\bf q}\cdot\dot{\bf q}}\, .
\end{equation}
For simplicity, we denote 
\begin{equation}
\xi_{i} = 2a_i\sin\psi \dot{q}_i+\dot{q}_i^2\, .
\end{equation}
After some calculations, we have that
\begin{equation}
\begin{split}
(vp_{1}\chi)^2 - (a_1\sin\psi)^2 =& (1 - v^{2}p^{2}_{1})\xi_{1} - v^{2}p^{2}_{1}\xi_{2}\, , \\   
(vp_{2}\chi)^2 - (a_2\sin\psi)^2 =& (1 - v^{2}p^{2}_{2})\xi_{2} - v^{2}p^{2}_{2}\xi_{1}\, . 
\end{split}
\end{equation}
As a consequence, we have the system given by
\begin{equation}
    \left[\begin{array}{cc}
1-v^{2}p^{2}_{1} &  -v^{2}p^{2}_{1} \\
 -v^{2}p^{2}_{2} & 1-v^{2}p^{2}_{2}
   \end{array}\right]  
   \left[\begin{array}{c}
\xi_1 \\
\xi_2
\end{array} \right]
   =
   \left[\begin{array}{c}
\chi^{2}v^{2}p_1^2 - a_1^2\sin^2\psi  \\
\chi^{2}v^{2}p_2^2 - a_2^2\sin^2\psi 
   \end{array} \right]\, ,
\label{eq:quadraticaXi}   
\end{equation}
which has solutions if and only if
\begin{equation}
    1-v^2({\bf p}\cdot{\bf p}) > 0,
\end{equation}
such condition is the same for the isotropic case. Under the condition we can solve eq.~(\ref{eq:quadraticaXi}) and through this solution we solve a quadratic system and we get
\begin{equation}
\dot{\bf q}=\frac{v\left(\cos \psi +{\bf h}\cdot \mathbf{q}\right){\bf p}}{\sqrt{1-v^{2}({\bf p}\cdot{\bf p})}}-\sin\psi{\bf a}_\phi\, .
\label{hamilt}
\end{equation}
Taking into account that the Hamiltonian is given by
\begin{equation}
{\cal H}_s(\mathbf{p},\mathbf{q})={\bf p}\cdot\dot{\bf q} -{\cal L}_s({\bf q},\dot{\bf q})\, ,   
\end{equation}
with ${\cal L}_s$ being the Lagrangian we have the main text expression.
\label{Apppendix:B}

\section{Hamiltonian's second-order derivatives}
\label{appendix:H2}

In order to prove the last equality of eq.~(\ref{eq:qdd}), we take eq.~(\ref{hamilton}), apply the second-order derivatives in $({\bf q},{\bf p})$ coordinates in the central-ray direction, we obtain the following equalities,
\begin{equation}
\begin{split}
\frac{\partial^2 {\cal H}_s}{\partial {\bf p}\partial {\bf p}^T} &= v_0\cos \psi_0{\bf I}\, ,\\ 
\frac{\partial^2 {\cal H}_s}{\partial {\bf q}\partial {\bf q}^T} &= \frac{\cos\psi_0}{v_0^2}\frac{\partial^2 v_0}{\partial{\bf q}\partial {\bf q}^T}\, , \\
\frac{\partial^2 {\cal H}_s}{\partial {\bf p}\partial {\bf q}^T} &= \frac{\partial^2 {\cal H}_s}{\partial {\bf q}\partial {\bf p}^T} = {\bf O}\, ,
\end{split}
\label{eq:derivada2pq}
\end{equation}
where ${\bf O}$ is the $2\times 2$ null matrix. Therefore, from eqs.~(\ref{sisthamilton}), (\ref{eq:sisthamilton0}), and~(\ref{eq:derivada2pq}), yields
\begin{equation}
\frac{d}{ds}\left(\frac{\partial{\cal H}_s}{\partial{\bf p}}\right) =  \frac{\partial^2 {\cal H}_s}{\partial {\bf p}\partial {\bf q}^T}\frac{d{\bf q}}{ds} + \frac{\partial^2 {\cal H}_s}{\partial {\bf p}\partial {\bf p}^T}\frac{d{\bf p}}{ds} = 0\, .  
\end{equation}

\section{Coordinate transformation }

In order to consider the forward and inverse transformations between Cartesian to ray-centered coordinates, we start with generalized-type momentum $\bar{\bf p}=[p_1,p_2,p_3]^T$, $t(\bar{\bf q})=\tau(\bar{\bf r}(\bar{\bf q}))$, and applying the chain rule for derivatives to $\tau$, yields
\begin{equation}
\bar{\bf p} =  \frac{\partial t}{\partial\bar{\bf q}} = \frac{\partial\bar{\bf r}^T}{\partial\bar{\bf q}}\frac{\partial\tau}{\partial\bar{\bf r}} = \bar{\bf E}\bar{\bf w}_q \, , 
\label{eq:pEg}
\end{equation}
where, using eq.~(\ref{vicinitypoint}), follows the expression
\begin{equation}
\bar{\bf E} = \frac{\partial\bar{\bf r}^T}{\partial\bar{\bf q}} = \left[\hat{\bf e}_1,\ \hat{\bf e}_2,\ \hat{\bf t} + ({\bf h}\cdot{\bf q})\hat{\bf n}\right]^T\, ,    
\end{equation}
which implies in
\begin{equation}
|\det\bar{\bf E}| = |\cos\psi + ({\bf h}\cdot{\bf q})|\, .   
\label{eq:detE}
\end{equation}

Eq.~(\ref{eq:pEg}) relates the displacement along a raypath with the slowness vector, $\bar{\bf w}_q$, related to the wavefront and  described, for ${\bf q}={\bf 0}$, by the expression
\begin{equation}
\bar{\bf w}_q = \frac{\partial\tau}{\partial\bar{\bf r}} = \frac{\partial\tau}{\partial\bar{\bf x}} = \bar{\bf w}\, .  
\end{equation}

The explicit expression of the coordinate transformation given by $\bar{\bf E}$ has advantages when we are interested in the description of the wavefront propagation phenomenon. For this reason, making the perturbation of the coordinate change matrix, we have
\begin{equation}
d\bar{\bf E} = \sum_{i=1}^{3}\left(\frac{\partial\bar{\bf E}}{\partial q_i}dq_i + \frac{\partial\bar{\bf E}}{\partial p_i}dp_i\right)\,.   
\end{equation}
However, as the same way in \cite{Iversen:2021}, for any value of $d\bar{\bf p}$ one can use the explicit form of $\bar{\bf E}$ to show that
\begin{equation}
\sum_{i=1}^{3}\frac{\partial\bar{\bf E}}{\partial p_i}dp_i  = \bar{\bf O}\, ,  
\end{equation}
which imply in $\bar{\bf E}(\bar{\bf q},\bar{\bf p}) = \bar{\bf E}(\bar{\bf q})$, where $\bar{\bf O}$ is the $3\times 3$ null matrix. Therefore, applying $\bar{\bf p}$-derivative in eq.~\ref{eq:pEg}, yields
\begin{equation}
\bar{\bf I} = \frac{\partial\bar{\bf w}^T}{\partial\bar{\bf p}}\bar{\bf E}^T\, ,    
\end{equation}
that implies in the inverse expression given by
\begin{equation}
\bar{\bf F} = \bar{\bf E}^{-T} = \frac{\partial\bar{\bf w}^T}{\partial\bar{\bf p}}\, .
\label{eq:EF}
\end{equation}
Besides, by definition, the coordinate transformation between $d\bar{\bf q}$ and $d\bar{\bf r}$ and, by previously construction, the coordinate transformation between $d\bar{\bf p}$ and $d\bar{\bf w}$ are given by, respectively,
\begin{equation}
d\bar{\bf r} = \bar{\bf E}^T d\bar{\bf q} \, , \quad \text{and} \quad d\bar{\bf w} = \bar{\bf F}^T d\bar{\bf p}\, . 
\label{eq:drEdq}
\end{equation}
To take the explicit expression of the inverse of $\bar{\bf E}$, no algebraic difficulties are demanded. However, the introduction of an auxiliary notation is necessary for a better understanding.

So with the help of the vectors $\bar{\bf f}_1$ and $\bar{\bf f}_2$, we take the inverse matrix $\bar{\bf F}$ below
\begin{equation}
\bar{\bf F} = \frac{1}{{\bf h}\cdot{\bf q} + \cos\psi}\left[({\bf h}\cdot{\bf q})\hat{\bf e}_1 - \bar{\bf f}_1,\ ({\bf h}\cdot{\bf q})\hat{\bf e}_2 - \bar{\bf f}_2,\ \hat{\bf n}\right]^T\, .    
\end{equation}

The geometric interpretation related to this notation can be seen in Figure \ref{fig:7}. 
We cannot confuse coordinate change matrices with the one that relates local coordinate systems given by the bases $\{ \hat{\bf e}_1, \hat{\bf e}_2, \hat{\bf t}\}$ and $\{\bar{\bf f}_1,\bar{\bf f}_2,\hat{\bf n}\}$ defined in the calculations.
By eqs.~(\ref{grupofase}) and~(\ref{tang00}), we take the group vector as
\begin{equation}
\hat{\bf t} = \frac{\cos\psi}{v}\left(\frac{1}{v}\frac{\partial v}{\partial\bar{\bf w}} + v^2\bar{\bf w}\right)\, ,   
\end{equation}
and substituting in eq.~\ref{vicinitypoint}, we obtain
\begin{equation}
\frac{d\bar{\bf r}}{ds} = \frac{\cos\psi}{v}\left(\frac{1}{v}\frac{\partial v}{\partial\bar{\bf w}} + v^2\bar{\bf w}\right) + ({\bf h}\cdot{\bf q})\hat{\bf n} +  \sum_{\text{i}=1}^2\hat{\bf e}_i\frac{dq_i}{ds}\, .
\end{equation}
From Equations~\ref{eq:drEdq} and~\ref{eq:EF}, we have the following relationship
\begin{equation}
{\bf F}\frac{d\bar{\bf r}}{ds} = \frac{d\bar{\bf q}}{ds}\, ,  
\label{eq:dqFdr}
\end{equation}
in other words, we can rewrite eq.~\ref{eq:dqFdr} in another way as follows
\begin{equation}
\begin{split}
\bar{\bf F}\frac{d\bar{\bf r}}{ds} &= \bar{\bf F}\left(\frac{\cos\psi}{v^2}\frac{\partial v}{\partial\bar{\bf w}} + (\cos\psi + {\bf h}\cdot{\bf q})\hat{\bf n} + \sum_{\text{i}=1}^2\hat{\bf e}_i\frac{dq_i}{ds}\right)\, ;  \\  
\frac{d\bar{\bf q}}{ds} &=  \left[\dot{\bf q} + \sin\psi{\bf a}_\phi,1\right] + \frac{\cos\psi}{v^2}\frac{\partial v}{\partial\bar{\bf p}}\, .   
\end{split}
\end{equation}
By definition, the variation of the space vector by the arc length is given by 
\begin{equation}
\frac{d\bar{\bf q}}{ds} = \left[\dot{\bf q}, 1\right]\, ,    
\end{equation}
consequently, we have the explicit relationship given by
\begin{equation}
\bar{\bf F}\frac{\partial v}{\partial\bar{\bf w}} = \frac{\partial v}{\partial\bar{\bf p}} = -v^2\tan\psi\left[{\bf a}_\phi^T, 0\right]^T\, .
\end{equation}
From the above expression, since the matrix $\bar{\bf F}$, has an inverse counterpart, it follows the closed form given by
\begin{equation}
\frac{1}{v^2}\frac{\partial v}{\partial\bar{\bf w}} = \left(\hat{\bf e}_1\cos\phi + \hat{\bf e}_2\sin\phi\right)\tan\psi\, ,    
\end{equation}
from which we take the well-defined expression for $\sin\psi{\bf a}_\phi$ given in the following
\begin{equation}
\frac{\cos\psi}{v^2}\frac{\partial v}{\partial{\bf p}} = -\sin\psi{\bf a}_\phi\, .
\label{eq:dvdp}
\end{equation}

Finally, examining equation \ref{eq:dvdp} in the central-ray direction, it follows 
\begin{equation}
\frac{\cos\psi_0}{v_0^2}\frac{\partial v_0}{\partial{\bf p}} = -\sin\psi_0{\bf a}_\phi\, ,
\label{crucial}
\end{equation}
from which we can summarize this section with two important notes. First, we can satisfactorily characterize the change of coordinates through the explicit matrices. Secondly, we can obtain an explicit expression for the ray-parameter $\sin\psi{\bf a}_\phi$. Note that the expression given by eq.~(\ref{eq:dvdp}) is fundamental to our goal of explicitly describing the ray's propagation.

\section{Discussion about variations on Ray-velocity vector}
\label{Ap:devP}

To compute the important second-order derivative of the group velocity vector, which is fundamental for modeling purposes. It is necessary to establish a relationship between the velocity variation and the direction regardless of the adopted coordinate system. Observe that the phase velocity squared is a second-order homogeneous function in the phase component. As a consequence, we can take the expression
\begin{equation}
\nu(\bar{\bf x},\bar{\bf w})^2 = v(\bar{\bf x},\|\bar{\bf w}\|\hat{\bf n}(\bar{\bf w}))^2 = \|\bar{\bf w}\|^2 v(\bar{\bf x},\hat{\bf n}(\bar{\bf w}))^2\, ,
\end{equation}
 thus, we can work with the Hamiltonian as
\begin{equation}
{\cal H}_t(\bar{\bf x},\bar{\bf w}) = \frac{1}{2}\nu(\bar{\bf x},\bar{\bf w})^2\, .  
\label{key}
\end{equation}
The first-order derivative in the phase parameter is obtained as
\begin{equation}
\frac{\partial{\cal H}_t}{\partial\bar{\bf w}} = \frac{1}{2}\frac{\partial\nu^2}{\partial\bar{\bf w}} = v^2\bar{\bf w} + \frac{1}{v}\frac{\partial v}{\partial\bar{\bf w}}\, ,     
\end{equation}
such an expression can be used for phenomenon modeling. To make practical such a procedure, from eq. ~\ref{key}, we have the following result
\begin{equation}
\frac{\partial^2{\cal H}_t}{\partial\bar{\bf w}\partial\bar{\bf w}^T} = \frac{1}{2}\frac{\partial^2\nu^2}{\partial\bar{\bf w}\partial\bar{\bf w}^T} = v^2\bar{\bf I} + \frac{\partial v^2}{\partial\bar{\bf w}}\otimes\bar{\bf w} + \frac{\partial}{\partial\bar{\bf w}}\left(\frac{1}{v}\frac{\partial v}{\partial\bar{\bf w}}\right)^T\, .
\label{metric}
\end{equation}
By using the previously introduced coordinate transformations, one can show that
\begin{equation}
\frac{\partial}{\partial p_j}\left(\frac{1}{v_0}\frac{\partial v_0}{\partial p_i}\right)^T = \frac{\bar{\bf f}_j}{\cos\psi}\cdot\left[\frac{\partial}{\partial\bar{\bf w}}\left(\frac{1}{v}\frac{\partial v}{\partial\bar{\bf w}}\right)^T\right]\frac{\bar{\bf f}_i}{\cos\psi}\, ,  
\label{tr1}
\end{equation}
and
\begin{equation}
\frac{\bar{\bf f}_j}{\cos\psi}\cdot\left[\frac{\partial v^2}{\partial\bar{\bf w}}\otimes\bar{\bf w}\right]\frac{\bar{\bf f}_i}{\cos\psi} = -2\left(\frac{v_0}{\cos\psi}\right)^2\left(\bar{\bf f}_j\cdot\hat{\bf n}\right)\left(\bar{\bf f}_i\cdot\hat{\bf n}\right)\, .  
\label{tr2}
\end{equation}
Applying the coordinate transformation on  ~\ref{metric} and collecting eqs.~(\ref{tr1}) and~(\ref{tr2}), it follows the relation given by eq.~(\ref{apEeq}) in the main text.

\end{document}